\documentclass[useAMS, usenatbib, usegraphicx]{mn2e} 

\title[Quasar outflows at high-$z$] 
{Quasar outflows at $z \geq 6$: the impact on the host galaxies} 

\author[P. Barai et al.] 
{Paramita Barai$^{1, 7}$, 
Simona Gallerani$^{1}$, 
Andrea Pallottini$^{1, 2, 3}$, 
Andrea Ferrara$^{1}$, 
\newauthor 
Alessandro Marconi$^{4, 5}$, 
Claudia Cicone$^{6}$, 
Roberto Maiolino$^{2, 3}$, 
Stefano Carniani$^{2, 3}$ 
\vspace{0.2cm} \\ 
$^{1}$ Scuola Normale Superiore, Piazza dei Cavalieri 7, I-56126 Pisa, Italy (E-mail: paramita.barai@sns.it) \\ 
$^{2}$ Cavendish Laboratory, University of Cambridge, 19 J. J. Thomson Ave., Cambridge CB3 0HE, UK \\ 
$^{3}$ Kavli Institute for Cosmology, University of Cambridge, Madingley Road, Cambridge CB3 0HA, UK \\ 
$^{4}$ Dipartimento di Fisica e Astronomia, Universita di Firenze, via G. Sansone 1, I-50019, Sesto Fiorentino (Firenze), Italy \\ 
$^{5}$ INAF - Osservatorio Astrofisico di Arcetri, Largo E. Fermi 5, I-50125, Firenze, Italy \\ 
$^{6}$ INAF - Osservatorio Astronomico di Brera, via Brera 28, I-20121, Milano, Italy \\ 
$^{7}$ Instituto de Astronomia, Geof\'isica e Ci\^encias Atmosf\'ericas (IAG-USP), Universidade de S\~ao Paulo, Brazil 
} 

\begin{document} 

\maketitle 


\begin{abstract} 

We investigate quasar outflows at $z \geq 6$ by performing zoom-in cosmological hydrodynamical simulations. 
By employing the SPH code GADGET-3, we zoom in the $2 R_{200}$ region around a 
$2 \times 10^{12} M_{\odot}$ halo at $z = 6$, inside a $(500 ~ {\rm Mpc})^3$ comoving volume. 
We compare the results of our {\it AGN} runs with a control simulation in which only stellar/SN feedback is considered. 
Seeding $10^5 M_{\odot}$ BHs at the centers of $10^{9} M_{\odot}$ halos, we find the following results. 
BHs accrete gas at the Eddington rate over $z = 9 - 6$. 
At $z = 6$, our most-massive BH has grown to $M_{\rm BH} = 4 \times 10^9 M_{\odot}$. 
Fast ($v_{r} > 1000$ km/s), powerful ($\dot{M}_{\rm out} \sim 2000 M_{\odot}$/yr) 
outflows of shock-heated low-density gas form at $z \sim 7$, and propagate up to hundreds kpc. 
Star-formation is quenched over $z = 8 - 6$, 
and the total SFR (SFR surface density near the galaxy center) is reduced by a factor of $5$ ($1000$). 
We analyse the relative contribution of multiple physical process: 
(i) disrupting cosmic filamentary cold gas inflows, (ii) reducing central gas density, (iii) ejecting gas outside the galaxy; 
and find that AGN feedback has the following effects at $z = 6$. 
The inflowing gas mass fraction is reduced by $\sim 12 \%$, 
the high-density gas fraction is lowered by $\sim 13 \%$, 
and $\sim 20 \%$ of the gas outflows at a speed larger than the escape velocity ($500$ km/s). 
We conclude that quasar-host galaxies at $z \geq 6$ are accreting non-negligible amount of cosmic gas, 
nevertheless AGN feedback quenches their star formation 
dominantly by powerful outflows ejecting gas out of the host galaxy halo. 

\end{abstract}







\section{Introduction} 
\label{sec-intro} 

Active galactic nuclei (AGN) are believed to host supermassive black holes (SMBHs) at their centers 
\citep[e.g.,][]{rees84, Kormendy95, Ferrarese05}. 
The enormous amounts of energy emitted by AGN is generated from accretion of matter onto the SMBHs 
\citep[e.g.,][]{Salpeter64, King03}. 
Quasars are very powerful AGN existing more commonly at high-$z$ than in the local Universe. 
Recent observational techniques have started to discover an increasing number of quasars at $z \geq 6$ 
\citep[e.g.,][]{Fan06, Mortlock11, Venemans13, Carnall15, Matsuoka16}. 
Far-infrared (IR) emission lines, which are unaffected by dust extinction, 
has lately been proven to be an important tool to probe $z \geq 6$ quasar host galaxies. 
In particular, the [CII] $158\mu$m fine structure line \citep[e.g.,][]{Maiolino05, Walter09}, 
and CO emission \citep[e.g.,][]{Riechers09, Wang10, Gallerani14, Stefan15}, 
from cool atomic and molecular gas \citep[review by][]{Carilli13, Gallerani17}, 
are often observed from these high-$z$ sources. 
The far-IR emission has been modelled in early $z \sim 6$ galaxies 
using cosmological hydrodynamical simulations \citep{Vallini15, Pallottini15, Pallottini16}. 

According to most theoretical and observational studies, 
feedback from AGN strongly influences the formation and evolution of galaxies, 
affecting the environment from pc to Mpc scales 
\citep[e.g.,][]{richstone98, Granato04, sazonov05, Barai08, Fabian12, Wagner13}. 
Cosmological simulations indicate that galaxies build up by the inflow of cosmic filamentary gas, 
a fraction of which accretes to the centers feeding and growing the central BHs, triggering an AGN phase. 
When the AGN is active for a period of time, 
its feedback energy output affects the hot/cold gas reservoirs in galaxies by 
expelling out and heating up some gas, as well as stopping gas inflows, thereby quenching star formation. 
This limits further gas accretion onto the SMBH, turning the AGN dead. 
Afterwards with the passage of time, gas might cool and inflow again, 
feeding the SMBH and triggering another phase of AGN. 
The result is a self-regulating process, where BHs and host galaxies coevolve during their simultaneous growth. 
The above processes are regarded to develop correlations in the central BH mass and properties 
(e.g. stellar mass, stellar velocity dispersion) of the host galaxy, 
that have been observed \citep[e.g.,][]{Magorrian98, Silk98, Gebhardt00, Shankar06}. 



A strong manifestation of AGN feedback are AGN outflows observed in a wide variety of forms 
\citep[reviews by][]{Crenshaw03, Everett07}. 
Some examples are: 
blue-shifted broad absorption lines in the ultraviolet and optical \citep{Reichard03, Rupke11}, 
warm absorbers \citep{chartas03, Krongold07} and ultra-fast outflows in X-rays \citep{Tombesi13, Tombesi15}, 
molecular outflows and atomic outflows detected in the IR, sub-millimetre and millimetre wavelengths 
\citep{Sturm11, Cicone14, Dasyra15, Feruglio15, Morganti16}, 
ionized gas in rest-frame optical \citep{Kakkad16}. 

Quasar outflows have also been observed in the early Universe. 
In the host galaxy of the quasar SDSS J1148+5251 at $z = 6.4$, 
\citet{Maiolino12} detected broad wings of the [CII] line tracing a massive outflow with velocities up to $\pm 1300$ km/s. 
Follow-up by \citet{Cicone15} inferred 
that the outflow has a complex morphology with the cold gas extended up to $30$ kpc, 
and revised the mass outflow rate lower limit to $1400 M_{\odot}$/yr. 
The physical mechanisms by which quasar outflows affect their host galaxies remain as open questions. 

SMBHs of mass $\geq 10^9 M_{\odot}$ are observed to be in place in luminous quasars by $z \sim 6$, 
when the Universe was less than $1$ Gyr old \citep[e.g.,][]{Willott03, Kurk07, DeRosa14, Wu15}. 
It is difficult to understand how these early SMBHs formed over such short time-scales, 
and there are open issues with various plausible scenarios 
\citep[e.g.,][]{Dijkstra08, Mayer10, Inayoshi12, Tanaka14, Matsumoto15}. 
The presence of massive BH seed candidates ($10^5 M_{\odot}$ at $z > 6$), possibly direct collapse black holes, 
in a bright Lyman-$\alpha$ emitter at $z = 6.6$ have been recently suggested \citep{Smith16, Pacucci17}. 

Concordance galaxy formation models based on cold dark-matter cosmology 
widely invoke AGN feedback as a crucial ingredient to self-regulate galaxy and SMBH growth. 
This has been studied in numerical hydrodynamical simulations 
\citep[e.g.,][]{DiMatteo05, Dubois10, Ostriker10, Barai11a, Hirschmann14}, 
as well as semi-analytical models \citep[e.g.,][]{Kauffmann00, Shankar04, Bower06, Somerville08}. 


AGN feedback should operate mostly in the negative form quenching star formation, 
as suggested by observations \citep[e.g.,][]{Schawinski06, Wang07, Lanz16}, 
and simulations \citep[e.g.,][]{Scannapieco05, vandeVoort11, Dubois13, Tremmel16}. 
This feedback limits the formation of massive stellar systems, enabling simulations to reproduce 
the observed exponential cut-off at the bright-end of galaxy luminosity function 
\citep[e.g.,][]{Croton06, Silk12}, as opposed to the dark-matter halo mass function. 
Populations of old, passive (the red and dead) massive elliptical galaxies are observed at $z \sim 2$ 
\citep[e.g.,][]{Cimatti04, Saracco05, Whitaker13}. 
This suggests that quasar feedback was already suppressing star formation at high-$z$. 

At the same time, AGN feedback can occasionally be positive, by inducing star-formation, 
and this aspect also plays an important role. 
AGN outflows can overpressure and compress clumpy gas clouds, triggering starbursts, 
as have been shown in theoretical and numerical studies \citep[e.g.,][]{DeYoung89, Silk05, Zubovas13}, 
including cold molecular clumps condensing out in quasar outflows \citep{Ferrara16}. 
Positive feedback has been observed in jet-induced star formation and radio-optical alignment 
\citep[e.g.,][]{Chambers87, Zinn13}, as well as in Seyfert-like radio-quiet AGN \citep{Cresci15}. 

AGN feedback in high-$z$ galaxies has been probed in a few previous numerical studies. 
Using zoom-in simulations, \citet{Costa14} examined 
the environment of bright quasars at $z \sim 6$ computing the thermal X-ray emission. 
Furthermore, \citet{Costa15} explored fast cold gas (originated by radiative cooling) 
moving together with the hot bipolar AGN outflows. 
\citet{Richardson16} performed zoom simulations of a forming galaxy cluster, 
and present AGN feedback results at $z = 5$, 
finding that AGN growth is self-regulated, regardless of Eulerian versus Lagrangian numerical methods used. 
\citet{Bieri16} executed radiation-hydrodynamical simulations of idealised gas-rich galaxy disks, 
where photons from a quasar interacts with the multiphase interstellar medium (ISM), 
and are found to generate powerful infrared-radiatively-driven outflows.
Using a $(400 h^{-1} {\rm Mpc})^3$ cosmological simulation, 
\citet{Waters16} predicted the properties of AGN populations at $z = 8 - 14$, 
and found $(10^6 - 10^8) M_{\odot}$ black holes accreting close to their Eddington luminosity. 

In this paper, we perform cosmological zoom-in simulations of quasar outflows at $z \geq 6$. 
Our goals are to investigate the impact of AGN outflows on host galaxies in the early Universe, 
and compute the outflow properties. 
We aim to determine the dominant physical mechanism(s) affecting star formation, 
by examining the relative contributions of: 
{\it (i)} halting of cold inflows, 
{\it (ii)} reducing central gas density and hence the mass that can be converted to stars, 
{\it (iii)} ejecting gas outside galaxy. 
In order to simulate massive quasar-host galaxies within the cosmological context, a large periodic boxsize is needed. 
At the same time, a low enough gas particle mass is desired, in order to resolve the smaller-scale structures. 
We decide to execute zoomed-in simulations to optimize the computational resource requirements, 
together with simulating a large cosmological box using the desired resolution. 

This paper is organised as follows: 
we describe our numerical code and simulation setup in \S\ref{sec-numerical}, 
present and analyze our results in \S\ref{sec-results}, 
and in \S\ref{sec-conclusion} we give a summary of our main findings. 


\section{Numerical Method} 
\label{sec-numerical} 

We use a modified version of the TreePM (particle mesh) - 
SPH (smoothed particle hydrodynamics) code {\sc GADGET-3} \citep{Springel05}. 
It includes an improved version of SPH \citep{Beck16} consisting of 
a higher-order Wendland C4 interpolating kernel, a time-dependent artificial viscosity term, 
and an artificial conduction term. 
Our simulations are outlined in \S\ref{sec-num-Sim}, and the different runs in \S\ref{sec-num-Runs}. 

The physical mechanisms of star-formation, supernovae (SN) explosions, 
gas accretion onto SMBHs, and resulting energy feedback are complex, 
with the gas motion driven by both thermal and radiation pressure. 
The relevant physical scales are orders of magnitude below the scales resolved in current galaxy formation simulations. 
Hence such physical processes are incorporated 
in the simulations using {\it sub-resolution} numerical prescriptions \citep[e.g.,][]{SDH05}. 
The physics of the gas is modelled using spatially averaged properties 
describing the medium on scales that are resolved in cosmological simulations. 
The sub-resolution physics that we use are described in \S\ref{sec-num-cool-SF-SN}, 
with the BH feedback prescriptions detailed in \S\ref{sec-num-BH-Accr-Feed}. 


\subsection{Simulations} 
\label{sec-num-Sim} 

We perform zoomed-in cosmological hydrodynamical simulations of high-redshift quasar-host galaxies. 
The initial conditions are generated using the 
{\sc MUSIC}\footnote{MUSIC - Multi-scale Initial Conditions for Cosmological Simulations: 
https://bitbucket.org/ohahn/music} software \citep{Hahn11}, following the steps below. 
A concordance flat $\Lambda$CDM model is used, with the cosmological parameters \citep[][results XIII]{Planck15}: 
$\Omega_{M, 0} = 0.3089, \Omega_{\Lambda, 0} = 0.6911, \Omega_{B, 0} = 0.0486, 
H_{0} = 67.74$ km s$^{-1}$ Mpc$^{-1}$. 
We express distances in the comoving scale unless differently stated. 

First, a dark-matter (DM) only low-resolution simulation is carried out 
of a $(500 ~ {\rm Mpc})^3$ comoving volume, using $256^3$ DM particles. 
Here the DM particle mass is $2 \times 10^{10} M_{\odot}$, 
and the gravitational softening length is $33 /h$ kpc comoving. 
The cosmological box is evolved from $z = 100$ up to $z = 6$, using periodic boundary conditions. 
Halos are identified within it using the {\it Friends-of-Friends} (FOF) group finder algorithm. 
We select the most-massive halo at $z = 6$ for the purpose of zoom-in. 
It has a total mass $M_{\rm halo} = 4.4 \times 10^{12} M_{\odot}$, and a virial radius $R_{200} \simeq 511$ kpc comoving. 
The halo mass and virial radius in comoving coordinates ($R_{200}$) are related 
such that $R_{200}$ encloses a density $200$ times the mean comoving matter density of the Universe: 
\begin{equation} 
\label{eq-Mhalo} 
M_{\rm halo} = \frac{4 \pi}{3} R_{200}^3 \left(200 \rho_{\rm crit} \Omega_{M,0}\right) , 
\end{equation} 
where $\rho_{\rm crit} = 3 H_0^2 / (8 \pi G)$ is the present critical density. 

We select the DM particles around the most-massive halo, 
belonging inside a cubic box of side $2 R_{200} \simeq 1022$ kpc comoving at $z = 6$. 
These DM particles are tracked back to our initial condition at $z=100$, 
and  the Lagrangian region occupied by them is determined. 
It is a volume of size $(5.21 ~ {\rm Mpc})^3$. 
The Lagrangian region is populated with particles of higher resolution: DM and baryons, 
going over $5$ further levels of refinement. 
While the coarser resolution has $8$ levels of refinement, the highest resolution has $13$ refinement levels. 

Finally, we restart a zoom-in simulation of the selected halo containing 
the high-resolution DM and gas particles $(N_{\rm part} = 2 \times 591408)$ in the central $(5.21 ~ {\rm Mpc})^3$ region, 
and the low-resolution $(N_{\rm part} = 17224370)$ DM particles 
outside the central zoomed region populating the remaining of the $(500 ~ {\rm Mpc})^3$ volume. 
The high-resolution particle masses are: $m_{\rm DM} = 7.54 \times 10^{6} M_{\odot}$, 
and $m_{\rm gas} = 1.41 \times 10^{6} M_{\odot}$. 
We employ $L_{\rm soft} = 1 /h$ kpc comoving as the Plummer-equivalent softening length for gravitational forces, 
for these high-resolution DM and gas particles. 

We have checked and ruled out any possible contamination by low-resolution particles 
within the zoomed volume during the simulations. 
There are no low-resolution DM particles inside a spherical region of radius 
$2 R_{200}$ around the most-massive galaxy center at each snapshot redshift. 
We verified this at all of our $37$ snapshots distributed between $z = 20$ and $z = 6$. 
Usually a contamination level of $< 1\%$ could be allowed in zoom-in simulations.

%

\begin{table*} 
\begin{minipage}{1.0 \linewidth} 
\caption{ 
Simulation runs and parameters. 
} 
\label{Table-Sims} 
\begin{tabular}{@{}cccccc} 

\hline 

Run & AGN feedback & Reposition of BH & Geometry of region where & Half opening angle & Total mass of zoomed-in \\ 
name & algorithm & to potential-minimum & feedback energy is distributed & of effective cone & halo at $z = 6$ \\ 

\hline 

{\it noAGN} & No BH & -- & -- & -- & $2.4 \times 10^{12} M_{\odot}$ \\   

{\it AGNoffset} & Kinetic & No & Bi-Cone & $45^{\circ}$ & $2.4 \times 10^{12} M_{\odot}$ \\   

{\it AGNcone} & Kinetic & Yes & Bi-Cone & $45^{\circ}$ & $1.6 \times 10^{12} M_{\odot}$ \\   

{\it AGNsphere} & Kinetic & Yes & Sphere & $90^{\circ}$ & $2.1 \times 10^{12} M_{\odot}$ \\   

\hline 
\end{tabular} 

\end{minipage} 
\end{table*} 


\subsection{Cooling, Star-Formation, SN Feedback}    
\label{sec-num-cool-SF-SN} 

Radiative cooling and heating is implemented by adopting the cooling rates 
from the tables of \citet{Wiersma09a}, which includes metal-line cooling. 
The photoionization code CLOUDY \citep[e.g.,][]{Ferland98} was used 
to pre-compute the cooling tables, taking into account the following. 
Eleven element species (H, He, C, Ca, O, N, Ne, Mg, S, Si, Fe) are tracked, 
and the gas is assumed to be dust-free, optically thin and in ionization equilibrium. 
Heating from a spatially-uniform time-dependent photoionizing radiation is considered 
from the cosmic microwave background and the ultraviolet/X-ray background (assuming the \citet{Haardt01} model). 

Star-formation is implemented following the multiphase effective sub-resolution model by \citet{SH03}. 
Gas particles with density above a limiting threshold, 
$n_{\rm SF} = 0.13$ cm$^{-3}$ (in units of number density of hydrogen atoms), represent regions of the ISM 
containing cold clouds in pressure equilibrium with hot gas. 
Collisionless star particles are spawned from these high-density gas particles, 
based on the stochastic scheme by \citet{Katz96}. 

Kinetic feedback from supernovae is included applying the energy-driven wind prescription. 
The wind mass-loss rate ($\dot{M}_{\rm SN}$) relates to the SF rate ($\dot{M}_{\star}$) as: 
$\dot{M}_{\rm SN} = \eta \dot{M}_{\star}$. 
We adopt a value for the wind mass loading factor, $\eta = 2$ \citep[e.g.,][]{Tornatore07, Tescari11, Barai13}, 
following observations revealing that SN-driven outflow rates in galaxies are comparable to 
or a few times larger than their SF rates \citep[e.g.,][]{Martin99, Pettini02, Bouche12, Newman12}. 
The wind kinetic energy is a fixed fraction $\chi$ of SN energy: 
$\frac{1}{2} \dot{M}_{\rm SN} v_{\rm SN}^2 = \chi \epsilon_{\rm SN} \dot{M}_{\star}$. 
Here $v_{\rm SN}$ is the wind velocity, and $\epsilon_{\rm SN} = 1.1 \times 10^{49}$ erg $M_{\odot}^{-1}$ 
is the average energy released by SN for each $M_{\odot}$ of stars formed. 
We adopt a constant-velocity outflow with $v_{\rm SN} = 350$ km/s \citep[as was done in e.g.][]{Barai15, Biffi16}. 

Stellar evolution and chemical enrichment are computed for the 11 elements \citep[following][]{Tornatore07}. 
Each star particle is treated as a simple stellar population (SSP). 
Given a stellar initial mass function (IMF), the mass of the SSP is varied in time 
following the death of stars, and accounting for stellar mass losses. 
We include a fixed stellar IMF from \citet{Chabrier03}, in the mass range $(0.1 - 100) M_{\odot}$. 
Stars within a mass interval $[8 - 40] M_{\odot}$ become SN first 
before turning into stellar-mass black holes at the end of their lives, 
while stars of mass $> 40 M_{\odot}$ are allowed to directly end in black holes without contributing to enrichment. 

Different yields are used from Type Ia SN \citep{Thielemann03}, Type II SN \citep{Woosley95}, 
and asymptotic giant branch stars \citep{vandenHoek97}. 
Stellar populations release metals with mass-dependent time delays, 
employing the lifetime function by \citet{Padovani93}. 
The mass range for SN-II is $M / M_{\odot} > 8$, 
while that for SN-Ia originating from binary systems is $0.8 < M / M_{\odot} < 8$ with a binary fraction of $10\%$. 
Both SN-Ia and SN-II also contribute to energy feedback within the chemical evolution model of \citet{Tornatore07}. 
In addition, there is mass loss through stellar winds and SN explosions, 
which are self-consistently computed for the given IMF and lifetime function. 
A fraction of a star particle's mass is restored as diffuse gas during its evolution, 
and distributed to the surrounding gas. 
The ejected energy (coupled thermally) and enriched material are spread 
among the neighbouring gas particles with weights given by the SPH kernel. 


\subsection{BH Models} 
\label{sec-num-BH-Accr-Feed} 

\subsubsection{BH Seeding} 

A BH (collisionless sink particle of mass $M_{\rm BH}$) is seeded at the center of each massive galaxy, 
whenever it reaches a total mass $M_{\rm halo} > 10^{9} M_{\odot}$, which does not contain a BH already. 
Halos are identified by executing a FOF group finder on-the-fly within our simulations. 
Galaxies are tracked simultaneously using the subhalo finder {\it SubFind}, 
which associates substructures to FOF halos. 
The center of each galaxy is considered as the location of the gravitational potential minimum of its subhalo. 

The seed BHs have an initial mass $M_{\rm BH} = 10^5 M_{\odot}$. 
Such massive seeds correspond to the direct collapse BH formation scenario. 

\subsubsection{BH Accretion} 

The BHs subsequently grow by accreting surrounding gas and by merger with other BHs. 
Gas accretion onto a BH is parametrized by the Bondi rate $(\dot{M}_{\rm Bondi})$ 
and is limited to the Eddington rate $(\dot{M}_{\rm Edd})$, 
\begin{equation} 
\label{eq-Mdot-BH} 
\dot{M}_{\rm BH} = {\rm min} \left( \dot{M}_{\rm Bondi}, \dot{M}_{\rm Edd} \right). 
\end{equation} 
Here, $\dot{M}_{\rm Bondi}$ is the Bondi-Hoyle-Lyttleton rate \citep{Hoyle39, Bondi44, Bondi52}: 
\begin{equation} 
\label{eq-Mdot-Bondi} 
\dot{M}_{\rm Bondi} = \alpha \frac{4 \pi G^2 M_{\rm BH}^2 \rho}{ \left(c_{s}^2 + v^2\right) ^ {3/2}} , 
\end{equation} 
where $G$ is the gravitational constant, $\rho$ is the gas density, $c_{s}$ is the sound speed, 
and $v$ is the velocity of the BH relative to the gas. 
We set $\alpha = 100$ as a numerical boost factor 
\citep[as done by e.g.,][]{SDH05, Khalatyan08, Johansson09a, Dubois13}. 
In Eq.~(\ref{eq-Mdot-BH}), 
$\dot{M}_{\rm Edd}$ is the Eddington mass accretion rate expressed in terms of the Eddington luminosity, 
\begin{equation} 
\label{eq-LEdd} 
L_{\rm Edd} = \frac{4 \pi G M_{\rm BH} m_p c} {\sigma_T} = \epsilon_r \dot{M}_{\rm Edd} c^2 , 
\end{equation} 
where $m_p$ is the mass of a proton, $c$ is the speed of light, 
and $\sigma_T$ is the Thomson scattering cross-section for an electron. 

BH accretion is numerically implemented in the {\sc GADGET-3} code using a stochastic methodology 
to {\it swallow} neighbouring gas particles \citep[originally from][]{SDH05}, as described in \citet{Barai14, Barai16}.

\subsubsection{BH Feedback} 

A fraction of the accreted rest-mass energy is radiated away by each BH. The radiation luminosity is, 
\begin{equation} 
\label{eq-Lr-BH} 
L_r = \epsilon_r \dot{M}_{\rm BH} c^2, 
\end{equation} 
with $\epsilon_r$ being the radiative efficiency. 
We adopt the mean value for radiatively efficient accretion onto a Schwarzschild BH \citep{Shakura73}: $\epsilon_r = 0.1$. 
A fraction $\epsilon_f$ (feedback efficiency) of this radiated energy is coupled to the surrounding gas as feedback energy: 
\begin{equation} 
\label{eq-Edot-Feed} 
\dot{E}_{\rm feed} = \epsilon_f L_r = \epsilon_f \epsilon_r \dot{M}_{\rm BH} c^2. 
\end{equation} 

The BH feedback energy is distributed in the {\it kinetic} form 
\citep[introduced in][as energy-driven wind]{Barai14, Barai16} only, and no thermal feedback is included. 
Surrounding gas is driven outward at a velocity $v_w$ and mass outflow rate $\dot{M}_w$. 
Given the energy-conservation equation, 
\begin{equation} 
\frac{1}{2} \dot{M}_w v_w^2 = \dot{E}_{\rm feed} = \epsilon_f \epsilon_r \dot{M}_{\rm BH} c^2 , 
\end{equation} 
the outflow rate can be expressed in terms of the BH accretion rate, 
\begin{equation} 
\label{eq-MdotW-EDW} 
\dot{M}_w = 2 \epsilon_f \epsilon_r \dot{M}_{\rm BH} \frac{c^2}{v_w^2} . 
\end{equation} 
For the two free parameters, we use the values: $\epsilon_f = 0.05$ \citep[e.g.,][to fit the correlation between 
central BH mass and host galaxy stellar velocity dispersion observed in the local Universe]{DiMatteo08, Rasia15}, 
and $v_w = 10,000$ km/s. 
The latter is motivated by typical AGN wind velocities seen in observations with 
a few $1000$ to $10000$ km/s \citep[e.g.,][]{Ramirez08, Perna15, Williams16}. 




The implementation of the BH sub-resolution models in the {\sc GADGET-3} code involves 
computing physical quantities by kernel-weighted smoothing over gas particles neighboring each BH. 
The relevant kernel size, or the BH smoothing length $h_{\rm BH}$, is determined at each timestep 
(analogous to finding gas particle smoothing length) by implicit solution of the equation, 
\begin{equation} 
\label{eq-BH-Smooth} 
\frac{4}{3} \pi h_{\rm BH}^3 \rho_{\rm BH} = M_{\rm ngb} , 
\end{equation} 
where $\rho_{\rm BH}$ is the kernel estimate of the gas density at the position of the BH, 
and $M_{\rm ngb}$ is the mass of $200$ neighboring gas particles (same number of neighbors as for SPH). 
The numerical value of $h_{\rm BH}$ is inversely proportional to the density of gas in the immediate environment of a BH. 
BHs that lie at galaxy centers have a smaller $h_{\rm BH}$ than off-center BHs, 
because gas density within a galaxy peaks at the center. 
In our simulations, $h_{\rm BH}$ typically lies between $(1 - 20) h^{-1}$ kpc, with an average value of $2 h^{-1}$ kpc. 
The minimum value of $h_{\rm BH}$ is comparable to the gravitational softening length of the gas particles. 
Further details of $h_{\rm BH}$ can be found in Appendix \ref{sec-app}. 

The kinetic feedback energy from each BH is distributed to the gas within a distance $h_{\rm BH}$, 
lying inside a volume of a pre-defined geometry: bi-cone, or sphere. 
For the bi-cone geometry, two conical volumes are defined with the BH at the apex, 
and the cones along two diametrically opposed directions. 
The slant height of each cone is $h_{\rm BH}$, and the half-opening angle is taken as $45^{\circ}$. 
The cone-axis direction is considered as fixed for each BH. 
A fixed direction is randomly assigned to a BH in our simulations during its seeding at the halo center. 
For the sphere geometry, the radius is $h_{\rm BH}$ 
(this case is equivalent to a bi-cone with a half-opening angle of $90^{\circ}$). 

Gas particles lying within the pre-defined geometry are tracked, 
and their total mass $M_{\rm gas}^{\rm vicinity}$ is computed. 
These gas particles are stochastically selected and kicked into AGN wind, by imparting a one-time $v_w$ velocity boost. 
We use a probabilistic criterion, similar to other sub-resolution prescriptions in {\sc GADGET-3}. 
The probability for $i$'th gas particle within the volume to be kicked is calculated as: 
\begin{equation} 
\label{eq-probKick} 
p_i = \frac{\dot{M}_w \Delta t} {M_{\rm gas}^{\rm vicinity}} , 
\end{equation} 
where $\Delta t$ is the timestep, and $\dot{M}_w$ is the mass outflow rate obtained from Eq.~(\ref{eq-MdotW-EDW}). 
At a given timestep, all the gas particles within the pre-defined geometry have the same probability to be ejected. 
The inverse proportionality of $p_i$ with $M_{\rm gas}^{\rm vicinity}$ ensures that 
the number of particles kicked does not depend on the geometry of the volume, but depends on $\dot{M}_w$ only. 
The quantity $\dot{M}_w \Delta t$ is the mass of gas to be kicked. 
The probability $p_i$ is constructed such that the available gas particles (total mass $M_{\rm gas}^{\rm vicinity}$ 
within pre-defined geometry) are sampled to reproduce kicking at the rate given by $\dot{M}_w$, on average. 

A random number $x_i$, uniformly distributed in the interval $[0, 1]$, is drawn and compared with $p_i$. 
For $x_i < p_i$, the gas particle is given an AGN wind kick, such that its new velocity becomes: 
\begin{equation} 
\label{eq-vNew} 
\vec{v}_{\rm new} = \vec{v}_{\rm old} + v_w \hat{n} . 
\end{equation} 
The kick direction $\hat{n}$ is set radially outward from the BH. 

In this work we consider only 2 geometries of the region where BH feedback energy is distributed: 
bi-conical, and spherical. 
This is to bracket the range of physical processes happening in the real Universe, 
where AGN feedback is observed to operate anisotropically in complex manners. 
For simplicity, we chose a specific bi-cone for each BH, 
which should capture the main difference between bi-conical and spherical cases. 
Another possibility is to consider variations of the direction (and opening angle) of the bi-conical feedback, 
depending on neighboring gas angular momentum or BH spin. 
This requires additional simulations and specific comparison with observations 
to distinguish between the physical scenarios, and forms prospects for future work.

\subsubsection{BH Reposition} 

We incorporate a scheme for BH repositioning or {\it pinning}, 
\citep[also done in e.g.,][]{SDH05, Sijacki07, Booth09, Ragone-Figueroa13, Wurster13, Schaye15}. 
It is often denoted in the literature as the {\it BH advection algorithm}. 
Each BH is repositioned manually at each time-step to the center (minimum gravitational potential location) of its host galaxy. 
This is done to correct for dynamical movements of 
BH particles wandering away from galaxy centers by numerical effects in SPH simulations. 
We perform one run without BH repositioning, and its implications are discussed in \S\ref{sec-results}. 

However the {\it BH advection algorithm} has its limitations. 
It fails to capture the Gyr timescale of sinking orbits for BHs during galaxy mergers \citep[e.g.,][]{Governato94}. 
It results in a nearly immediate SMBH merger, as well as high BH accretion rates during merger events. 
As an improvement, \citet{Tremmel15, Tremmel16} proposed a sub-grid force correction term 
to model the dynamical friction of a SMBH as it orbits within its host galaxy. 
This new approach allows for long-lived dual SMBH orbits during a galaxy merger, 
and accurately follows a SMBH's gradual orbital decay. 


\subsubsection{BH Merging} 
\label{sec-num-BH-Merge} 

We assume that central BHs merge when their host galaxies merge during hierarchical structure formation. 
When two BH particles come near such that the distance between them is smaller than the smoothing length of either one, 
and their relative velocity is below the local sound speed, they are allowed to merge to form a single BH 
\citep[e.g.,][]{Sijacki07, DiMatteo12}. 


\subsection{Series of Simulation Runs} 
\label{sec-num-Runs} 

We execute a series of four simulations, with characteristics listed in Table~\ref{Table-Sims}. 
All the four runs incorporate metal cooling, chemical enrichment, SF and SN feedback. 
The first run has no AGN included, while the latter three explore different AGN feedback models: 

\begin{itemize} 

\item {\it noAGN} $-$ no BH present. This is a control simulation in which 
only cooling, enrichment, star-formation, and SN feedback are implemented. 

\item {\it AGNoffset} $-$ with AGN accretion, growth and feedback. 
Kinetic BH feedback distributed inside bi-cone ($45^{\circ}$ half opening angle). 

\item {\it AGNcone} $-$ Repositioning of BH to halo center. 
Kinetic feedback distributed in bi-cone ($45^{\circ}$ half opening angle). 

\item {\it AGNsphere} $-$ Repositioning of BH. 
Kinetic feedback distributed in sphere ($90^{\circ}$ half opening angle). 

\end{itemize} 

The zoomed-in halo evolves differently in the four runs because of the different feedback models. 
The precise value of total halo mass ($M_{\rm halo}$) at $z = 6$ is mentioned in the last column of Table~\ref{Table-Sims}.

\section{Results and Discussion} 
\label{sec-results}

\subsection{Black Hole Environment} 
\label{sec-res-BH-Environ} 

\begin{figure*} 
\centering 
\includegraphics[width = 1.05 \linewidth]{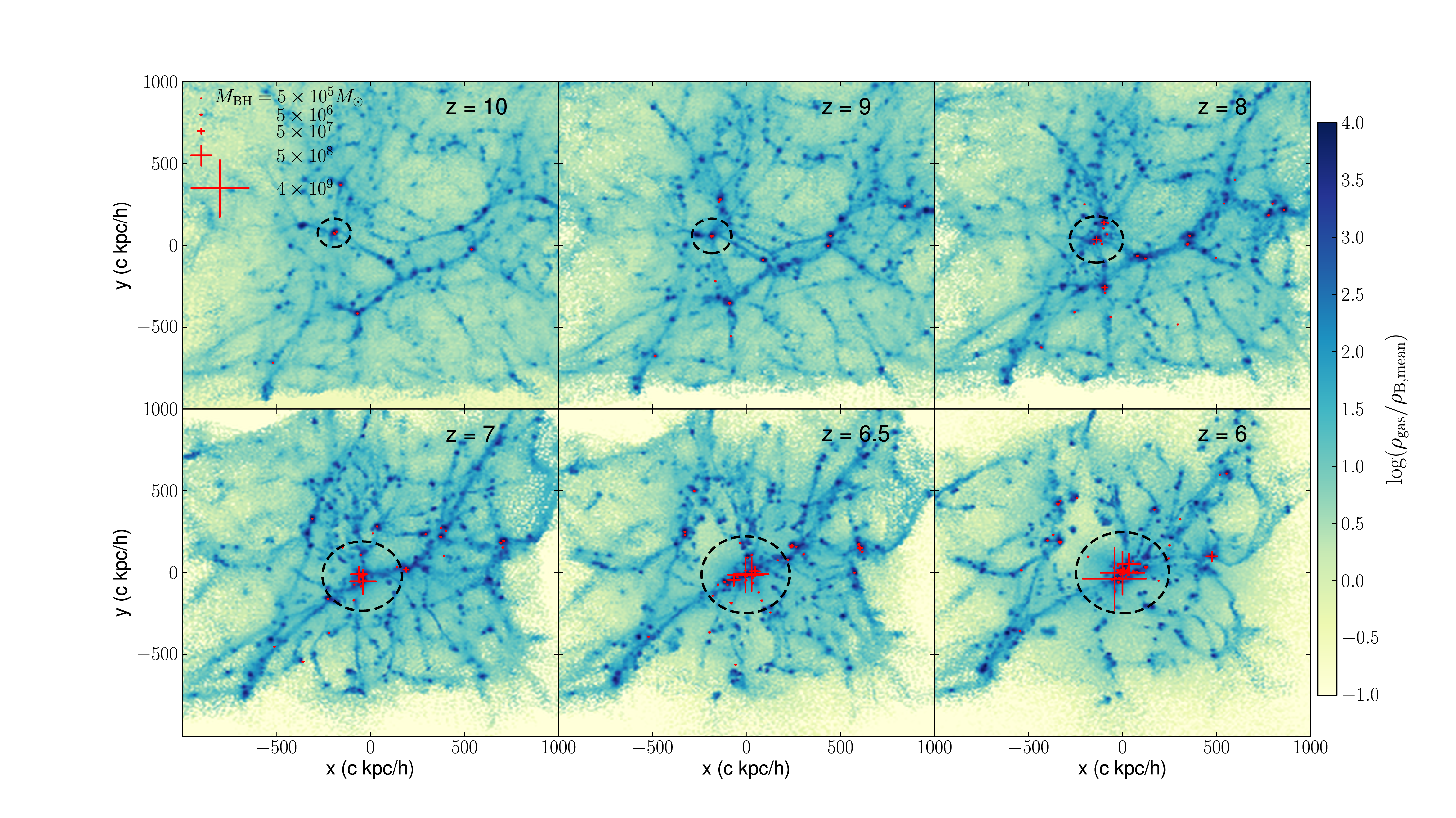} 
\caption{ 
Gas overdensity in the run {\it AGNcone} at six epochs $z = 10, 9, 8, 7, 6.5, 6$. 
Each panel shows a projected $(1000 h^{-1}$ kpc$)^3$ comoving volume 
around the location of the most-massive galaxy at $z = 6$. 
The red plus symbols designate positions of BHs, 
with the symbol size proportional to BH mass as indicated in the legend of top-left panel. 
The largest symbol (bottom-right panel) corresponds to $M_{\rm BH} = 4 \times 10^9 M_{\odot}$, 
and the smallest symbols (every panel) correspond to $M_{\rm BH} = 5 \times 10^5 M_{\odot}$. 
The black dashed circle is the virial radius $R_{\rm 200}$ of the most-massive galaxy at each epoch plotted. 
} 
\label{fig-GasOden-BHpos} 
\end{figure*} 

The spatial locations of the BHs within our {\it AGNcone} simulation box 
can be visualized in Fig.~\ref{fig-GasOden-BHpos}, overplotted with the gas overdensity. 
The plotted region is centered on the most-massive galaxy at $z = 6$. 
The six panels present six redshifts, each one showing a projected $(2000$ kpc/h$)^3$ comoving volume. 
The red plus symbols designate BH positions, where the symbol size is proportional to BH mass going from 
$M_{\rm BH} = 5 \times 10^5 M_{\odot}$ (smallest symbols in every panel) 
to $4 \times 10^9 M_{\odot}$ (largest symbol in the bottom-right panel). 
The black dashed circle is the virial radius $R_{\rm 200}$ of the most-massive galaxy at each epoch plotted. 

The earlier epochs ($z = 10, 9, 8$) show BHs being seeded at the centers of galaxies 
with $M_{\rm halo} > 10^{9} M_{\odot}$. 
These massive BH-host galaxies lie at the high-density intersections of cosmological large-scale-structure filaments. 
When two massive galaxies merge during hierarchical structure formation, 
their central BHs merge as well (according to the prescription in \S\ref{sec-num-BH-Merge}) to form a single larger BH. 
In addition the BHs grow in mass by accreting gas from their surroundings, 
as galaxies evolve and gas inflows to their centers. 
The later epochs ($z = 6.5, 6$) display a few BHs which have grown supermassive 
located near the center of the plotted region. 

Initially, the most-massive BH is seeded and grows in the most-massive galaxy, 
as seen in the top panels of Fig.~\ref{fig-GasOden-BHpos} $(z = 10, 9, 8)$. 
At the later epochs ($z = 7, 6.5, 6$, bottom panels), 
there is an offset between the most-massive galaxy center and the most-massive BH location. 
This is because of the difference of the rates at which a BH and its host galaxy grows in mass. 
Neighboring galaxies evolve at different rates, 
and the galaxy halo which grows to become most massive in the simulation is not the one 
hosting the most-massive BH, but somewhat offset from it. 

We find that in our simulations the BHs form a SMBH binary or triplet (in this run {\it AGNcone}) at the halo center, 
and form a merging system at $z = 6$. 
This is a physically plausible scenario since binary SMBHs are observed in the local Universe 
\citep[e.g.,][]{Ju13, Andrade-Santos16}. 
Recent studies have started to observe that quasars often occur in merging systems at high-$z$. 
E.g. at $z=4.8$, $50\%$ of quasar host galaxies ($3$ out of $6$) 
analysed by \citet{Trakhtenbrot17} are found to have companion sub-millimetre galaxies. 

However binary SMBHs at $z \geq 6$ have not been detected so far. 
This discrepancy could be because of limitations of current observations; 
for example angular resolution being too low and not sufficient to distinguish multiple sources, 
or BHs being highly obscured and not visible at optical/ultraviolet wavelengths. 
Thus our prediction that SMBHs may occur as binaries at $z = 6 - 7$ 
makes the case stronger for future observational programs to detect them. 


\subsection{Black Hole Accretion and Growth} 
\label{sec-res-BH-Growth} 

\begin{figure*} 
\centering 
\includegraphics[width = 1.1 \linewidth]{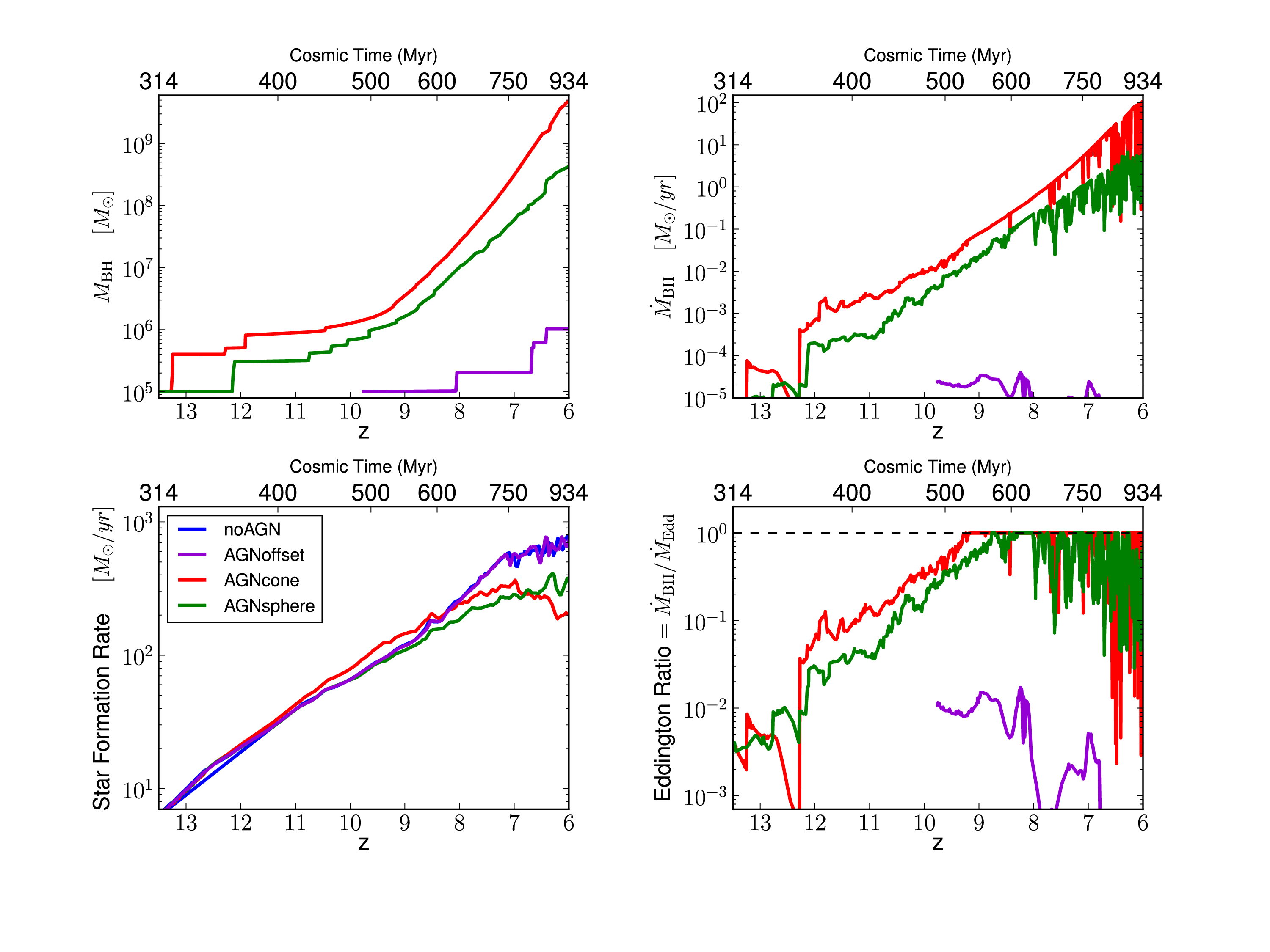} 
\vspace{-1.7cm} \\ 
\caption{ 
Evolution with redshift of BH mass (top-left panel), mass accretion rate (top-right), 
and Eddington ratio (bottom-right), of the most massive BH in each run. 
The star formation rate (total in whole simulation box) is plotted in the bottom-left panel. 
The different colors discriminate the runs as labelled in the bottom-left. 
} 
\label{fig-BH-Mass-AccrRate-SFRD} 
\end{figure*} 

We find that first BHs are seeded at $z \sim 15$ in our simulations, 
when the first halos reach $M_{\rm halo} = 10^{9} M_{\odot}$. 
In the runs {\it AGNcone} and {\it AGNsphere}, one of these first seeds grow to become the most-massive BH. 
However in run {\it AGNoffset}, the BH which becomes most-massive is seeded at $z \sim 10$. 
This variance in the seed epochs is because of the different BH growth modes, as described next. 

The redshift evolution of the most-massive BH in the three AGN runs is plotted in Fig.~\ref{fig-BH-Mass-AccrRate-SFRD}: 
BH mass in the top-left panel, BH mass accretion rate ($\dot{M}_{\rm BH}$) at the top-right, 
and Eddington ratio $= \dot{M}_{\rm BH} / \dot{M}_{\rm Edd}$ at the bottom-right panel. 
Each BH starts from an initial seed of $M_{\rm BH} = 10^5 M_{\odot}$, 
at $z \sim 14$ in the runs {\it AGNcone} and {\it AGNsphere} ($z \sim 10$ in {\it AGNoffset}). 
The subsequent growth is due to merger with other BHs (revealed as vertical rises in $M_{\rm BH}$), 
and gas accretion (visualized as the positive-sloped regions of the $M_{\rm BH}$ versus $z$ curve). 

The dominant mode of BH growth occurs over the redshift range $z = 9 - 6$ in runs {\it AGNcone} and {\it AGNsphere}, 
corresponding to Eddington-limited gas accretion where Eddington ratio $= 1$. 
The $\dot{M}_{\rm BH}$ has a power-law increase, and the BH mass increases by a factor $\sim 10^3$. 
The final properties reached at $z = 6$ depends on the simulation: 
$M_{\rm BH} = 4 \times 10^9 M_{\odot}$ and $\dot{M}_{\rm BH} = 100 M_{\odot}$/yr in run {\it AGNcone} (red curve), 
$M_{\rm BH} = 4 \times 10^8 M_{\odot}$ and $\dot{M}_{\rm BH} = 6 M_{\odot}$/yr in run {\it AGNsphere} (green curve), 
$M_{\rm BH} = 10^6 M_{\odot}$ and $\dot{M}_{\rm BH} < 10^{-5} M_{\odot}$/yr in run {\it AGNoffset} (violet curve). 
There is variability of the $\dot{M}_{\rm BH}$, whereby it fluctuates by a factor of up to $100$. 

The BH grows 10 times more massive at $z = 6$ in the {\it AGNcone} case than in the {\it AGNsphere} run. 
This is because more gas can inflow along the perpendicular direction to the bi-cone, and accrete onto the BH. 
As discussed later in \S\ref{sec-res-outflow}, the relative direction of the cosmic gas feeding the galaxy 
(and consequently its central BH) from the large-scale structures remains the same in all the runs. 
In the {\it AGNcone} case, a strong outflow develops perpendicular to this direction 
of cosmic gas infall (Fig.~\ref{fig-Gas-rho-T-SFR-velR}). 
Thus more gas from the cosmic large-scale structures continue to infall onto the BH and causes it to grow larger. 

Our results of early SMBH growth are consistent with previous zoom-in cosmological simulations of high-$z$ quasars. 
\citet{Sijacki09} found that, starting from seeds of $10^5 M_{\odot}$ around $z = 15$ 
in dark matter haloes of mass $10^9 - 10^{10} M_{\odot}$, 
it is possible to build up SMBHs of $10^9 M_{\odot}$ by $z = 6$ that assemble most of their mass during 
extended Eddington-limited accretion periods. 
\citet{Costa14} identified that the above is true only in the most-massive haloes located in the most-overdense regions. 
We zoomed-in just the most-massive halo from our parent $(500 ~ {\rm Mpc})^3$ box, 
where we find BHs growing to $\sim 10^9 M_{\odot}$ by $z = 6$. 
However, our BHs undergo Eddington-limited accretion up to later epochs ($z = 6$) as compared to \citet{Costa14}, 
who analysed that accretion becomes limited by AGN feedback by $z \sim 9 - 8$. 

We find that the BHs become supermassive ($M_{\rm BH} \sim 10^9 M_{\odot}$) 
in the simulations ({\it AGNcone} and {\it AGNsphere}) with BH repositioning, 
where a BH is repositioned to the center of its host galaxy. 
Instead in the case without BH repositioning (run {\it AGNoffset}, violet curve), 
the most-massive BH grows up to $10^6 M_{\odot}$ only, and the growth is dominated by mergers, 
since gas accretion is always occurring at low Eddington ratios $(\dot{M}_{\rm BH} / \dot{M}_{\rm Edd} \leq 0.02)$. 
This is because with no repositioning algorithm, a BH is offset from the center of its host galaxy. 
Hence the gas density in the BH vicinity 
is lower than the case where a BH lies at the galaxy center with a high gas density. 
This causes a smaller Bondi accretion rate $\dot{M}_{\rm Bondi}$ (see Eq.~\ref{eq-Mdot-Bondi}), 
resulting in a tiny BH growth. 
Therefore implementing BH repositioning is necessary for our BHs to encounter galaxy-central high-density gas, 
and accrete at the Eddington rate in order to reach $10^9 M_{\odot}$ by $z = 6$. 



\subsection{Star Formation} 
\label{sec-res-SF} 

Stars form in the simulation volume from cold dense gas. 
The star formation rate (total in the whole simulation box) versus redshift of the four simulations 
is displayed in Fig.~\ref{fig-BH-Mass-AccrRate-SFRD}, bottom-left panel. 
The SFR rises with time in all the runs initially, and continues to increase in the {\it noAGN} case without a BH. 
The SFR in run {\it AGNoffset} is almost similar to that in the run {\it noAGN}, 
because the BHs are too small there to generate enough feedback. 
A similar outcome happens in the runs {\it AGNcone} and {\it AGNsphere} at $z \geq 8$, when the BHs are too small. 

The star-formation mostly occurs over an extended region at galaxy centers, 
where cosmic large-scale-structure gas inflows and cools. 
The presence of a BH quenches star formation. 
We quantify and discuss later in \S\ref{sec-res-Quench} the details of the SF quenching mechanisms. 

The models suppress SF substantially from $z \sim 8$ onwards, 
when the BHs have grown massive and generate larger feedback energy. 
Thus, we find that BHs need to grow to $M_{\rm BH} > 10^7 M_{\odot}$, 
in order to suppress star-formation, even in massive galaxies 
(of $M_{\star} = 4 \times 10^{10} M_{\odot}$, and specific-SFR $= 5 \times 10^{-9}$ yr$^{-1}$). 
BH feedback causes a reduction of SFR up to $4$ times at $z = 6$: 
from $800 M_{\odot}$/yr in the {\it noAGN} run, 
to $200 M_{\odot}$/yr in run {\it AGNcone}, and $350 M_{\odot}$/yr in run {\it AGNsphere}. 



\subsection{Black Hole - Galaxy Correlation} 
\label{sec-res-BH-Gal-CoEvol} 

\begin{figure*} 
\centering 
\includegraphics[width = 1.15 \linewidth]{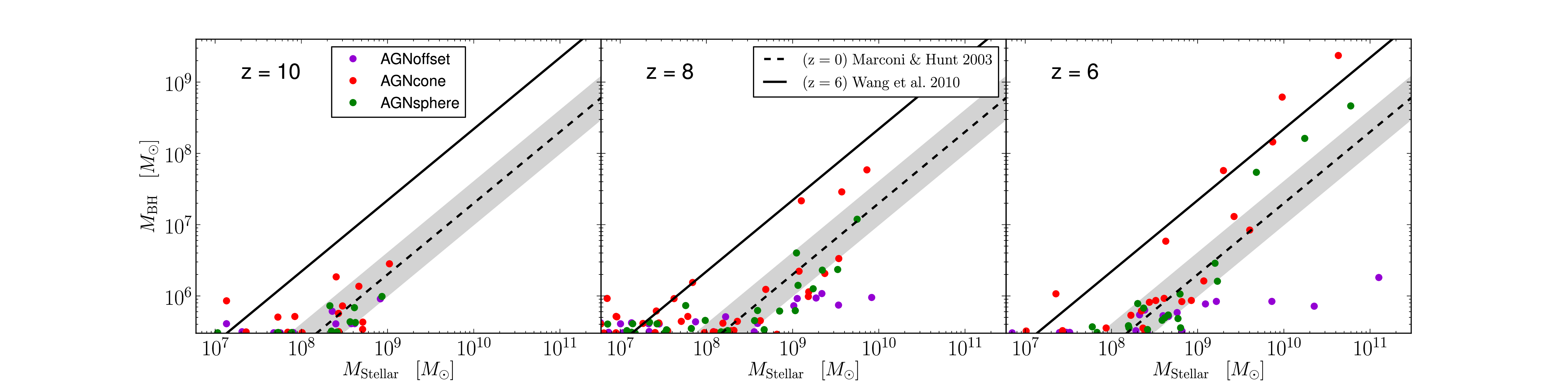} 
\caption{ 
BH mass versus stellar mass of all the galaxies within the zoomed-in volume, 
at three epochs $z = 10, 8, 6$, in the panels from left. 
The plotting color distinguish results from different runs: {\it AGNoffset} - violet, {\it AGNcone} - red, {\it AGNsphere} - green. 
The black lines indicate the observed BH mass versus stellar bulge mass relation of: 
local galaxies \citep{Marconi03} as the dashed line, and $z \sim 6$ quasars \citep{Wang10} as the solid line. 
} 
\label{fig-Mass-BH-vs-Stellar} 
\end{figure*} 

\begin{figure*} 
\centering 
\includegraphics[width = 1.1 \linewidth]{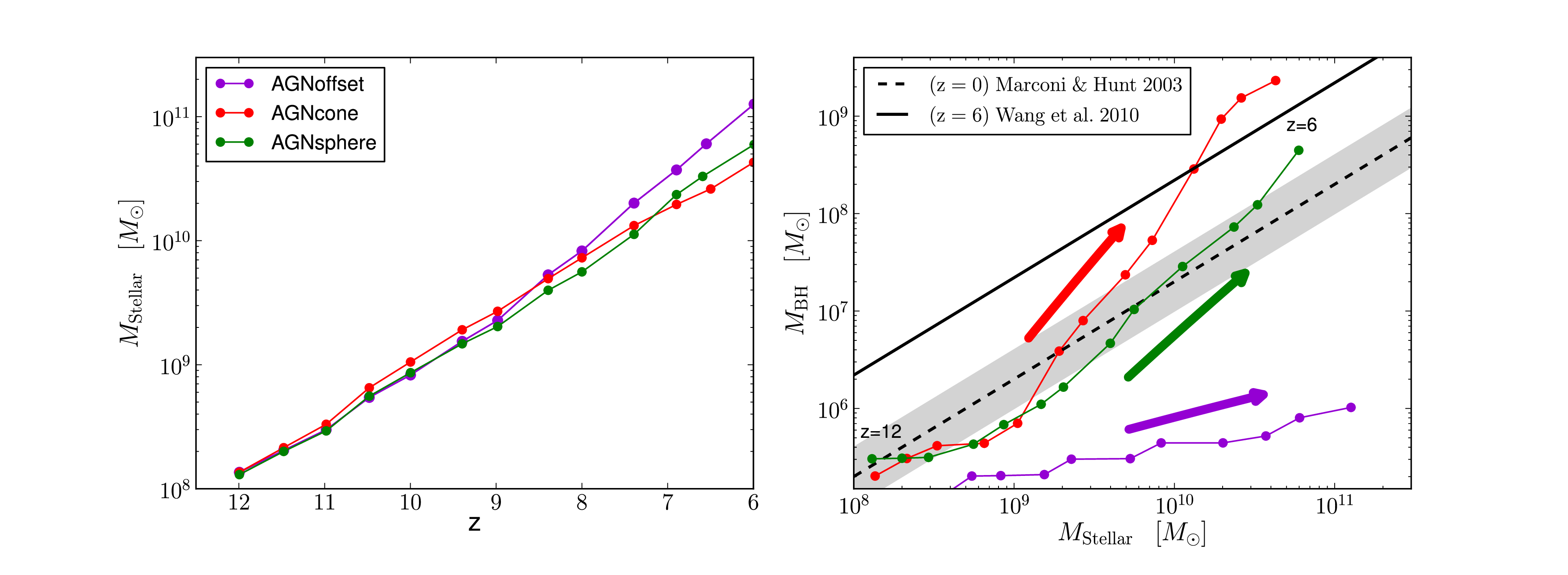} 
\caption{ 
{\it Left panel:} 
Redshift evolution of the stellar mass of the host galaxy of the most-massive BH progenitor. 
The plotting color denote different runs: {\it AGNoffset} - violet, {\it AGNcone} - red, {\it AGNsphere} - green. 
{\it Right panel:} Redshift track of the BH mass versus host galaxy stellar mass of the most-massive BH progenitor 
(which was plotted in Fig.~\ref{fig-BH-Mass-AccrRate-SFRD}). 
The thick arrows denote the direction of time evolution (going from high to low $z$) for the three runs. 
The starting $(z = 12)$ and ending $(z = 6)$ redshifts are written near the relevant points on the tracks. 
The black lines indicate the observed BH mass versus stellar bulge mass relation of: 
local galaxies \citep{Marconi03} as the dashed line, and $z \sim 6$ quasars \citep{Wang10} as the solid line. 
} 
\label{fig-Mass-Stellar-and-BH} 
\end{figure*} 

The BH - galaxy correlation obtained in our simulations is presented 
in Fig.~\ref{fig-Mass-BH-vs-Stellar} as the $M_{\rm BH}$ versus $M_{\star}$ (stellar mass) diagram. 
We define galaxy stellar mass as the mass of all star particles inside the subhalos obtained by the 
subhalo finder {\it SubFind} (our halo tracking algorithm has been described in \S\ref{sec-num-BH-Accr-Feed}). 
It shows all the galaxies within the zoomed-in volume, at three epochs $z = 10, 8, 6$, in the panels from left. 
The plotting color distinguishes results from different runs: 
{\it AGNoffset} - violet, {\it AGNcone} - red, {\it AGNsphere} - green. 
Observational data is overplotted as the black lines 
indicating the BH mass versus stellar bulge mass relationships at different epochs. 
Local galaxies ($z = 0$) fit the black-dashed line: $M_{\rm BH} / M_{\star} = 0.002$ \citep{Marconi03}. 
The ratio is observed to be steeper at high-$z$. 
Far-IR and CO bright $z \sim 6$ quasars lie along the black-solid line: 
median $M_{\rm BH} / M_{\star} = 0.022$ \citep{Wang10}. 
In our run {\it AGNoffset}, where the most-massive BH reaches $M_{\rm BH} \sim 10^6 M_{\odot}$ only, 
the $[M_{\rm BH} - M_{\star}]$ correlation is not reproduced. 

Our simulations (runs {\it AGNcone} and {\it AGNsphere}) show that, 
at $z \sim 6$ massive galaxies $(M_{\star} > 10^{9} M_{\odot})$ 
contain SMBHs $(M_{\rm BH} > 10^7 M_{\odot})$ more massive than expected from the local relation. 
This suggests that SMBHs grow faster than their host galaxies in the early Universe. 
Observations inferring the $[M_{\rm BH} - M_{\star}]$ correlation in ultraluminous 
$z \sim 6$ quasars \citep[e.g.,][]{Walter04, Venemans16, Wang16} also points to the same scenario. 
Though it should be noted that, at high-$z$ there are 
large uncertainties in quantifying the host galaxy stellar masses \citep[e.g.,][]{Valiante14}, 
which is usually estimated using the dynamical gas mass measured from CO and [CII] observations. 

We also see a dependence on the BH mass: less-massive BHs lie on the local relation, albeit with a large scatter; 
while they {\it migrate} to the $z = 6$ correlation as they grow. 
We find the limiting mass as $M_{\rm BH} \sim 10^7 M_{\odot}$, 
below which our results more closely match the observed $[M_{\rm BH} - M_{\star}]$ relation at $z = 0$. 
While at $M_{\rm BH} > 10^7 M_{\odot}$, 
the simulated galaxies more closely match the observed correlation at $z = 6$. 
Such a trend is consistent with observations \citep[e.g.,][]{Willott15, Wang16}, 
where however the limiting BH mass is $10$ times higher: 
most of the $z \sim 6$ quasars with $M_{\rm BH} \sim 10^8 M_{\odot}$ are close to the local relationship, 
while more-massive quasars tend to lie above. 

The right panel of Fig.~\ref{fig-Mass-Stellar-and-BH} shows the redshift track of the 
BH mass versus host galaxy stellar mass of the most-massive BH progenitor 
(which was plotted in Fig.~\ref{fig-BH-Mass-AccrRate-SFRD}). 
For each curve, the redshifts start from $z = 12$ (bottom-left point), and go up to $z = 6$ (top-left point). 
The red and green curves indicate that as BHs grow from $10^7 M_{\odot}$ to $10^8 M_{\odot}$, 
they migrate from the $z = 0$ to the $z = 6$ correlation. 
The corresponding redshift evolution of the host galaxy stellar mass of the same BH 
is shown in the left panel of Fig.~\ref{fig-Mass-Stellar-and-BH}. 


\subsection{Gas Outflows} 
\label{sec-res-outflow} 

While the BHs output feedback energy (\S\ref{sec-num-BH-Accr-Feed}), 
high-velocity gas propagates radially outward and shocks with the surrounding slower-moving gas. 
This creates bubble-like gas outflows originating from the central BH.

\subsubsection{Outflow Morphology} 
\label{sec-res-outflow-Morph} 

\begin{figure*} 
\centering 
\includegraphics[width = 0.97 \linewidth]{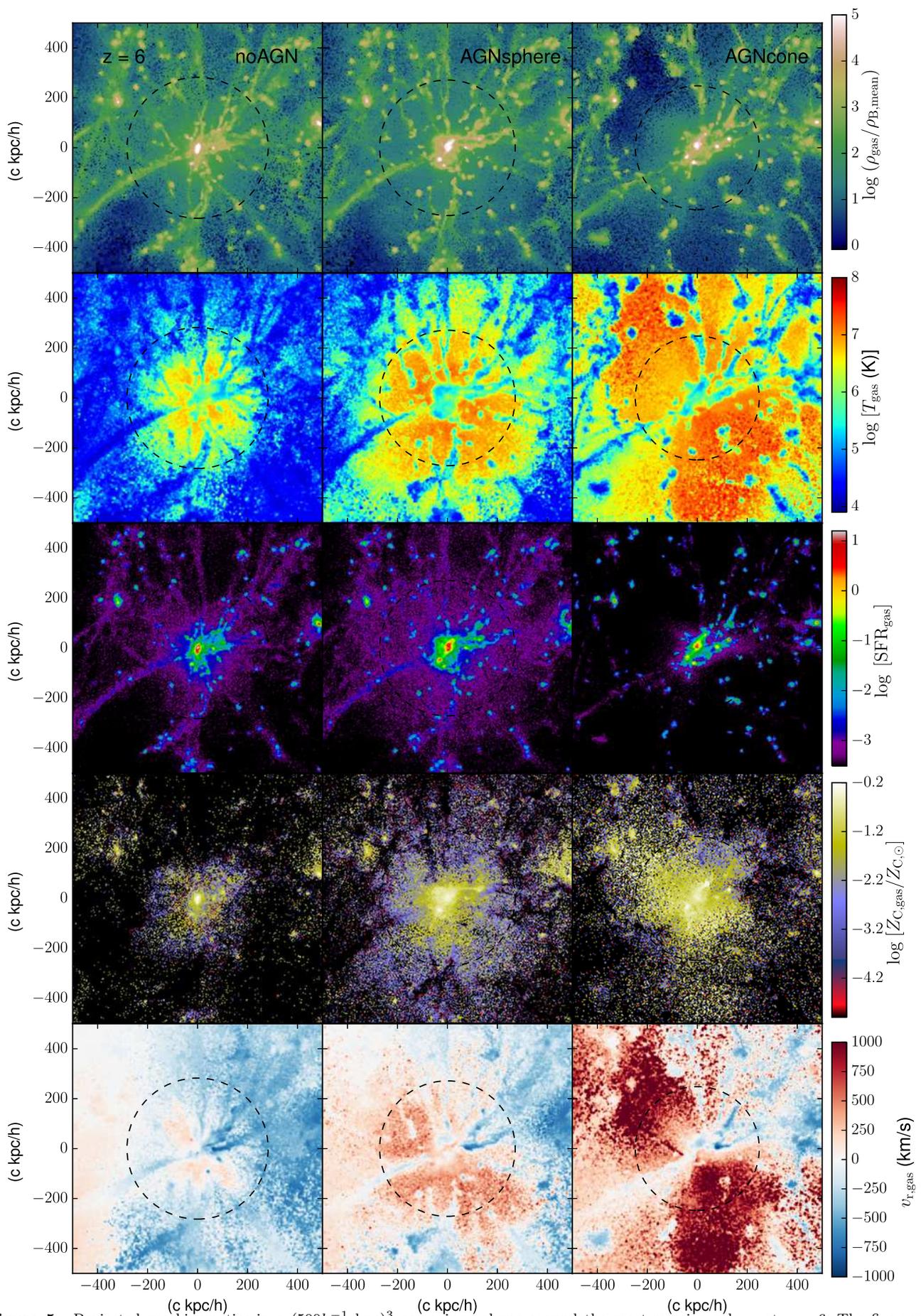} 
\caption{ 
Projected gas kinematics in a $(500 h^{-1}$ kpc$)^3$ comoving volume around the most-massive galaxy at $z = 6$. 
The five rows indicate gas properties: 
overdensity (first row -- top), temperature (second), SFR (third), carbon abundance (fourth), and radial velocity (fifth -- bottom). 
The three columns are different simulations: {\it noAGN} (left), {\it AGNsphere} (middle), and {\it AGNcone} (right). 
The black dashed circle depicts the galaxy virial radius $R_{\rm 200}$ in each case. 
} 
\label{fig-Gas-rho-T-SFR-velR} 
\end{figure*} 


The outflow morphology is plotted in Fig.~\ref{fig-Gas-rho-T-SFR-velR}, 
which displays the projected gas kinematics in a $(1000 h^{-1}$ kpc$)^3$ comoving volume 
in three runs, at $z = 6$. 
The overdensity, temperature, SFR, carbon abundance $(Z_{\rm C})$, and radial velocity $(v_{r})$ 
of the gas is plotted in the five rows from the top. 
We analyze the abundance of carbon, which is one of the most-abundant heavy element in the Universe. 
Observationally, the [CII] line can be used to characterize the ISM of galaxies, 
and CO emission can be used to track molecular outflows. 
We compute $Z_{\rm C}$ as the ratio of carbon mass to the total gas mass for each gas particle. 
Abundance ratios are expressed in terms of the Solar value: 
$Z_{{\rm C}, \odot} = 0.00218$ \citep[mass fraction of carbon in the Sun;][]{Asplund05}. 

In the {\it noAGN} run (left column), weak outflows ($v_{r} < 300$ km/s) develop 
bounded within $0.5 R_{\rm 200}$ as warm-hot ($T \sim 10^6$ K) halo gas. 
It is caused by SN feedback and galaxy merger shocks. 
The $Z_{\rm C}$ distribution is more centrally concentrated in this case. 
Such a high-mass $(M_{\star} \sim 10^{11} M_{\odot})$ 
galaxy cannot efficiently drive outflows with only SN feedback \citep[e.g.,][]{Benson03, Talia16}. 

The other two simulations ({\it AGNcone} and {\it AGNsphere}) show the formation of BH feedback-induced outflows, 
which are hot ($T \sim 10^8$ K) - visible as yellow and red areas in the temperature plot, 
and consist of low-density, metal-enriched gas. 
The outflows are fastest ($v_{r} > 2000$ km/s) in run {\it AGNcone} 
(right column, where the BHs become more-massive than the other runs), extended bipolar shaped, 
propagating to beyond the galaxy $R_{\rm 200}$ (black dashed circle). 
The outflows reach $v_{r} \sim 1500$ km/s in run {\it AGNsphere} (middle column), 
are more spherical shaped and limited within $R_{\rm 200}$. 
They disrupt the cold dense filamentary gas inflows to the galaxy center, along the direction of outflow propagation. 
This quenches star formation, and halts the formation of nearby satellite galaxies. 
It is revealed by the disrupted clumps and lack of star-forming dense filaments 
near the central galaxy in runs {\it AGNcone} and {\it AGNsphere}, 
as compared to run {\it noAGN} (which we quantify later in \S\ref{sec-res-Quench}). 
The outflows transport metals away from star-forming regions, 
and enrich the surrounding circumgalactic medium out to $R_{\rm 200}$. 
Some inflows of cold, dense gas continue to occur perpendicular to the outflow direction. 

We find that the density increment, at the edges of the outflow shocks, 
remains below the SF threshold density $n_{\rm SF}$. 
Therefore no new star-formation is triggered in our simulations by AGN feedback. 
This might be a numerical resolution effect, 
since our resolution (length scale of $1$ kpc comoving) is lower than those 
where positive feedback from AGN is simulated in galaxies \citep[e.g.,][]{Bieri15}.

\subsubsection{Gas Radial Velocity} 
\label{sec-res-RadVel} 

\begin{figure*} 
\centering 
\includegraphics[width = 1.15 \linewidth]{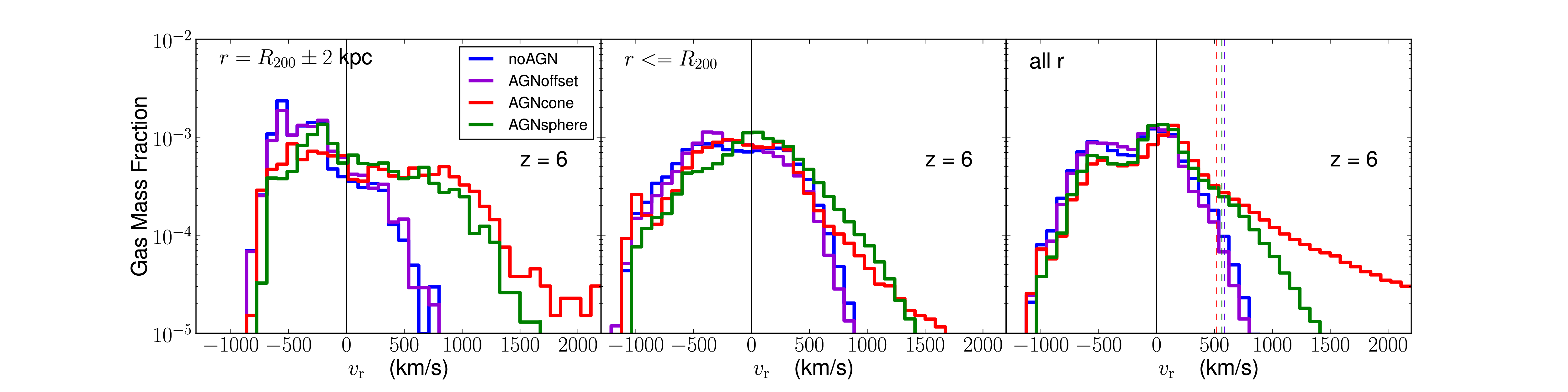} 
\caption{
Radial velocity histogram of gas within given regions around the most-massive galaxy at $z = 6$, in four simulations: 
{\it noAGN} - blue curve, {\it AGNoffset} - violet, {\it AGNcone} - red, {\it AGNsphere} - green. 
The regions are: gas at $r = R_{200} \pm 2$ kpc (left panel), gas inside $r <= R_{200}$ (middle), 
and all the gas within simulation volume (right). 
Each curve shows the mass fraction of gas per velocity bin, 
and the histogram is normalized to the total gas mass within the plotted region. 
The vertical black solid line is $v_r = 0$. 
The colored dashed lines in the right panel indicate the halo escape velocity in the four runs. 
} 
\label{fig-VelR-Hist-z6} 
\end{figure*} 

The radial velocity ($v_r$) histogram of gas at $z = 6$ is plotted in Fig.~\ref{fig-VelR-Hist-z6}. 
The three panels display gas within distinct regions: at $r = R_{200} \pm 2$ kpc comoving (left), 
inside $r <= R_{200}$ (middle), and all the gas within simulation volume (right). 
Each curve shows the mass fraction of gas per velocity bin, 
and the histogram is normalized to the total gas mass within the plotted region. 
The colored dashed lines in the right panel indicate the halo escape velocity in each run, 
$v_{\rm esc} = \sqrt{2 G M_{\rm halo} / R_{200}}$ (for galaxy total halo mass $M_{\rm halo}$). 

The $v_r$ distribution in the run {\it AGNoffset} (violet curve) is almost coincident with that in {\it noAGN} (blue), 
because there the BHs do not exert enough feedback. 
The velocities are the lowest in these two cases, reaching a maximum $v_r \sim 800$ km/s within the virial radius. 

We find a larger fraction of outflowing ($v_r > 0$) and a smaller fraction of inflowing ($v_r < 0$) gas 
in runs {\it AGNcone} (red) and {\it AGNsphere} (green), 
when compared to {\it noAGN}; which we quantify in \S\ref{sec-res-Frac-InOut}. 
Powerful high-velocity outflows are created by AGN feedback in these two runs, where the BHs grow supermassive. 
Gas reaches a maximum $v_r \sim 1400$ km/s in {\it AGNsphere}, and $v_r > 2000$ km/s in {\it AGNcone}. 
Our simulation velocity values are consistent with the observations of $z > 6$ quasar outflows 
by \citet{Maiolino12} and \citet{Cicone15}, 
where the gas traced by [CII] is moving at high velocities, up to $\sim 1400$ km/s. 
However note that the bulk of our simulated outflows consist of shock-heated gas of $T > 10^{6}$ K. 
While [CII] observations probe cold gas of a few $100$s K. 

Our $v_r$ values ($1400 - 2000$ km/s) also agree well with previous zoom-in simulation studies of quasar outflows. 
\citet{Costa14} found that AGN-driven outflows are highly anisotropic, 
pushing gas at  $\geq 1000$ km/s out to tens of kpc. 
\citet{Costa15} deduced that parts of hot AGN outflows, 
propagating through a clumpy medium pre-enriched with metals from SN, 
can cool radiatively, resulting in velocity widths of cold gas up to $\sim 2000$ km/s. 


\subsubsection{Outflow and Inflow Fractions of Gas} 
\label{sec-res-Frac-InOut} 

\begin{figure*} 
\centering 
\includegraphics[width = 1.1 \linewidth]{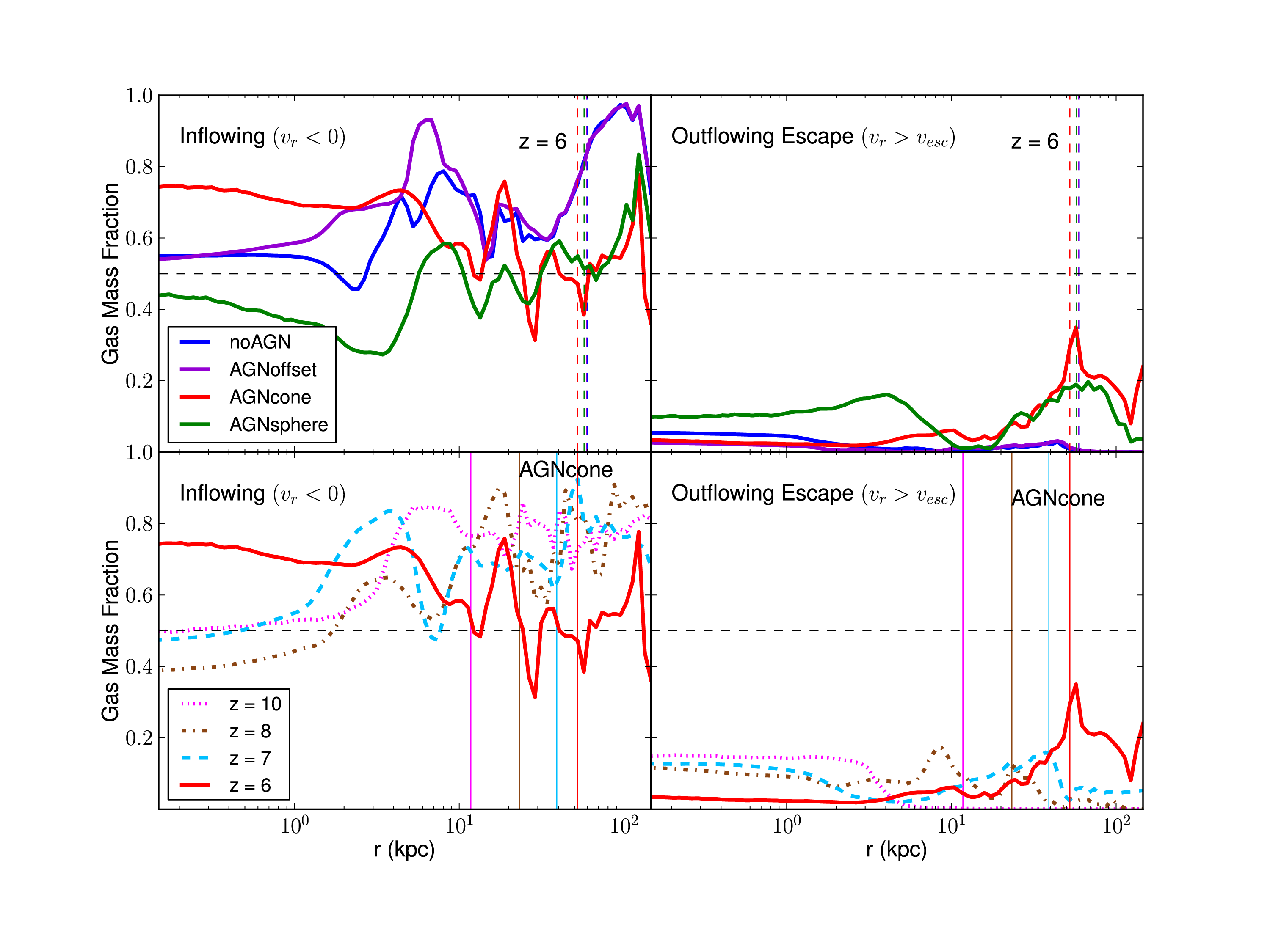} 
\caption{ 
Gas mass fraction inflowing ($v_r < 0$) in the left column, 
and outflowing at above the escape velocity ($v_r > v_{\rm esc}$) at the right column, 
versus radius of the most-massive galaxy in the simulations. 
Top row displays results at $z = 6$ of four runs: 
{\it noAGN} - blue, {\it AGNoffset} - violet, {\it AGNcone} - red, {\it AGNsphere} - green. 
Bottom row is at four redshifts of a single run {\it AGNcone}: 
$z = 10$ - magenta dotted, $z = 8$ - brown dot-dashed, $z = 7$ - deepskyblue dashed, and $z = 6$ - red solid. 
Each curve shows the in/out-flowing mass as a fraction of gas contained in the radial bins. 
The vertical colored lines (dashed in the top panels, and solid in the bottom) 
indicate the halo virial radius $R_{200}$ in a respective run at the relevant redshift. 
Note that the radial coordinate (x-axis) is in units of physical kpc. 
} 
\label{fig-Frac-InOut-Profile} 
\end{figure*} 

The radial\footnote{The radius is computed as the distance from the gravitational potential minimum 
of the galaxy, which is an output of the {\it SubFind} subhalo finder.} 
distribution of inflow and relevant outflow gas fractions is plotted in Fig.~\ref{fig-Frac-InOut-Profile}. 
The inflowing ($v_r < 0$) mass fraction is in the left column, 
which quantifies the inflow of cosmic gas and satellite galaxies. 
Gas fraction whose radial velocity is positive and exceeds $v_{\rm esc}$ ($v_r > v_{\rm esc}$) is at the right column, 
which indicates outflowing gas that can escape the halo potential. 

The top row of Fig.~\ref{fig-Frac-InOut-Profile} presents the results of four runs at $z = 6$. 
Near and beyond the virial radius $R_{200}$, the inflow mass fraction 
in the run {\it AGNoffset} (violet curve) is same as that in {\it noAGN} (blue), because there the BHs are too small. 
Whereas BH feedback induced outflows are created in the runs {\it AGNcone} (red) and {\it AGNsphere} (green): 
the inflow fraction around $R_{200}$ is lower by $30 \%$ than in the {\it noAGN} run, 
and consequently the outflowing escape fraction is higher by $30 \%$. 
The peaks in the inflow fraction curve correspond to accreting satellite galaxies and clumps, 
which occurs at different radii for each run. 
As an example, the inflow fraction in the inner regions at $r < 5$ kpc 
is the highest in run {\it AGNcone} ($\sim 70 \%$) compared to all other cases. 
This is because the AGN feedback process changes the merger history in the inner parts of the galaxy. 
The time taken by inner satellites to merge is larger in the presence of AGN feedback 
(also visible as the density substructures in the top panels of Fig.~\ref{fig-Gas-rho-T-SFR-velR}), 
because of disruption by powerful outflows. 

The bottom row of Fig.~\ref{fig-Frac-InOut-Profile} denotes the fractions at four redshifts in the run {\it AGNcone}, 
which visualizes the time evolution of the impact of AGN feedback. 
The inflow fractions near $R_{200}$ remain almost similar from $z = 10$ to $z = 7$. 
Feedback from the BH has a visible impact at $z = 6$, 
when the inflow fraction reduces by $30 \%$ at the galaxy outskirts ($r \sim 20 - 200$ kpc), 
and a similar fraction escapes the halo while outflowing. 

We find that BH outflows suppress cosmic gas inflows by $30 \%$ near $R_{200}$ at $z = 6$, 
and causes $(20 - 30) \%$ of the gas to reach velocities in excess of the halo escape speed. 
It also alters the merger history and substructure distribution in the inner parts of the galaxy.

\subsubsection{Mass Outflow Rate} 
\label{sec-res-MdotOut} 

\begin{figure*} 
\centering 
\includegraphics[width = 1.15 \linewidth]{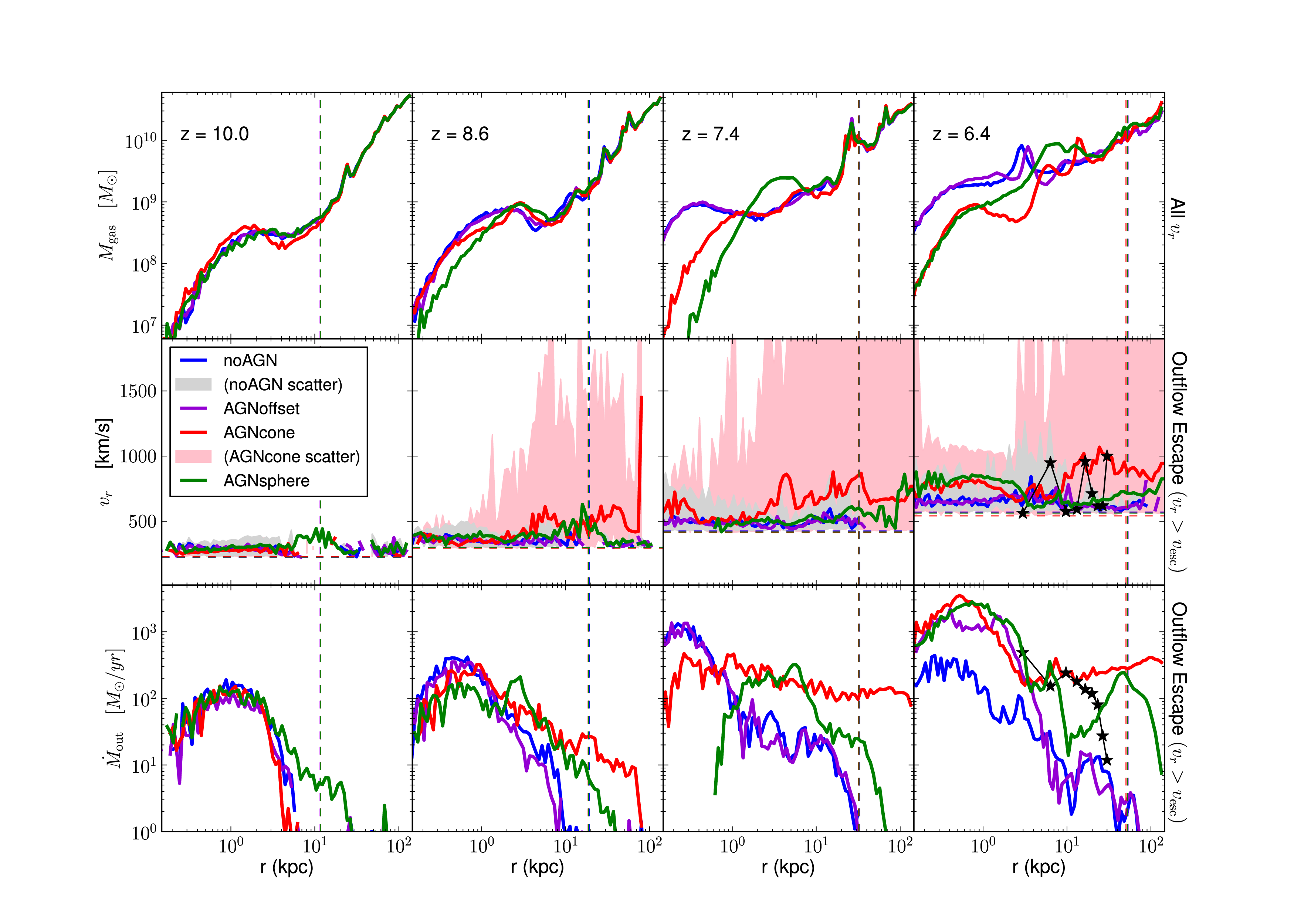} 
\caption{ 
Radial distributions of gas around the most-massive galaxy in the simulations: 
top row shows the total (any $v_r$) gas mass. 
The next two rows show properties of only the gas outflowing at above the escape velocity ($v_r > v_{\rm esc}$): 
median outflow radial velocity in the middle row, and mass outflow rate in the bottom row. 
Four redshifts are plotted in the columns from left: $z = 10, 8.6, 7.4, 6.4$. 
In each panel, the colors distinguish four runs: 
{\it noAGN} - blue, {\it AGNoffset} - violet, {\it AGNcone} - red, {\it AGNsphere} - green; 
with the vertical colored dashed lines indicating the halo virial radius ($R_{200}$). 
The radial scatter of the $v_r$ distribution is displayed in the middle row for two runs: 
{\it noAGN} - grey shaded area, {\it AGNcone} - pink shaded area; 
showing the $95$th percentiles above and below the median-$v_r$. 
Observational data of outflow velocity and mass-loss-rate versus distance \citep{Cicone15} 
at $z = 6.4$ are overplotted in the middle-right and bottom-right panels as the black star symbols. 
Note that the radial coordinate (x-axis) is in units of physical kpc. 
} 
\label{fig-Outflow-M-vr-Mdot-Profile} 
\end{figure*} 


We measure outflowing gas using spherical shell volumes around galaxies. 
Gas particles (the $i$'th particle having a mass $m_i$) lying inside a shell, 
at radius $r$ of thickness ${\Delta r}$, are selected. 
We calculate the mass outflow rate by summing over the particles inside ${\Delta r}$ 
having a radial velocity larger than the escape speed $(v_r > v_{\rm esc})$: 
\begin{equation} 
\label{eq-Mdot-Sph} 
\dot{M}_{\rm out} (r) = \sum_{v_{r, i} > v_{\rm esc}} \frac{m_i |v_{r, i}|} {\Delta r}. 
\end{equation} 

We plot in Fig.~\ref{fig-Outflow-M-vr-Mdot-Profile} the gas mass (top row), outflow velocity (middle row) 
and $\dot{M}_{\rm out}$ (bottom row) versus radius, at four redshifts (in the four columns). 
Note that the radial coordinate (x-axis) is in units of physical kpc in this figure. 
The profiles are the same for all the four runs at $z = 10$ (left column), 
when the BHs are too small $(M_{\rm BH} <\sim 10^6 M_{\odot})$ to produce a dynamically relevant influence. 
AGN feedback effects start to be seen at $z = 8.6$ (second column) in the run {\it AGNsphere} (green curve), 
which also suppressed star-formation from similar epochs (see Fig.~\ref{fig-BH-Mass-AccrRate-SFRD} bottom-left panel). 

The total gas mass is computed as: $M_{\rm gas} = \sum_{{\rm all}-v_{r, i}} m_i$. 
At $z \geq 8.6$ (three right columns), $M_{\rm gas}$ inside $2$ physical kpc 
is up to $10$ times lower in runs {\it AGNcone} and {\it AGNsphere}, compared to the {\it noAGN} case. 
This is because AGN feedback ejects central gas and reduces the inner gas content. 

The next two rows of Fig.~\ref{fig-Outflow-M-vr-Mdot-Profile} show properties of 
only the gas outflowing at above the escape velocity ($v_r > v_{\rm esc}$). 
The outflow radial velocity is in the middle row, where the solid curves depict the median-$v_r$. 
The radial scatter of the $v_r$ distribution is displayed for two runs as the shaded areas, 
showing the $95$th percentiles above and below the median-value. 
The ejection of outflows from the center are visible as the rise in the scatter of run {\it AGNcone} (pink shaded area), 
with some gas reaching $v_r > 2000$ km/s at $z = 7.4$ and $6.4$. 
Whereas in the {\it noAGN} run, the scatter (grey shaded area) is always small, limited to $v_r \leq 800$ km/s. 
The mass outflow rate is in the bottom row. 
In the outer regions $r \geq 1$ kpc, the $\dot{M}_{\rm out}$ is higher in the strong AGN feedback runs 
({\it AGNcone} and {\it AGNsphere}) at $z \geq 8.6$ (three right columns). 


Observational data of outflow velocity and mass-loss-rate versus distance \citep[Figure 5 and 8 of][]{Cicone15} 
at $z = 6.4$ are overplotted in the middle-right and bottom-right panels as the black asterisk symbols. 
The $v_{\rm out} \sim 1000$ km/s observed points agree well with the median-$v_r$ from our simulation {\it AGNcone}. 
Such extended ($r > 8 - 10$ kpc) high-velocity outflows cannot be explained with SF and SN-driven feedback only 
\citep[e.g.,][]{Valiante12}. 

In the observed radii range, $r = 3 - 30$ kpc, 
the $\dot{M}_{\rm out}$ in our strong AGN feedback runs is $10 - 100$ times higher than the {\it noAGN} case. 
The observational $\dot{M}_{\rm out}$ between $r = 3 - 15$ kpc agree well with our results in the {\it AGNcone} run. 
However, in the outer regions $r = 15 - 60$ kpc our simulations predict an almost flat $\dot{M}_{\rm out}$ versus $r$, 
while the observations show a steep decline of $\dot{M}_{\rm out}$ with $r$. 
This discrepancy is partly because of the different method used to measure the outflow rate in \citet{Cicone15}. 
The radial distance of the outflowing clouds from the galaxy center is used 
in the denominator of Eq.~(\ref{eq-Mdot-Sph}), i.e. $\dot{M}_{\rm out} (r) = \sum \frac{m_i |v_{r, i}|} {r_i}$. 
There is also the contribution of observational uncertainties. 
The decreasing $\dot{M}_{\rm out}$ with radius at $r \geq (20 - 30)$ kpc may be in part due to 
instrument sensitivity issues, and/or the outflowing clouds being 
ablated by dynamical instabilities or photo-evaporated \citep[for a discussion see][]{Ferrara16}. 
We finally note that for a fair comparison, the simulations must be processed to extract mock spectra; 
and $\dot{M}_{\rm out}$ computed henceforth using $r$ exactly like the way done in observational analyses. 

The observed integrated outflow rate obtained by adding up the mass-loss-rate contributions 
from all the outflowing clumps is $1400 \pm 300 M_{\odot}$/yr \citep{Cicone15}. 
Using the same $9$ bins as the observations, we obtain 
an integrated mass outflow rate of $128 M_{\odot}$/yr in run {\it noAGN}, and $2067 M_{\odot}$/yr in run {\it AGNcone}. 
Thus we conclude that the high outflow rate detected in observations are driven by the central quasar; 
and not by star-formation activities.

\subsection{Radial Profiles of Gas Properties} 
\label{sec-res-RadProfiles} 

\begin{figure*} 
\centering 
\includegraphics[width = 1.17 \linewidth]{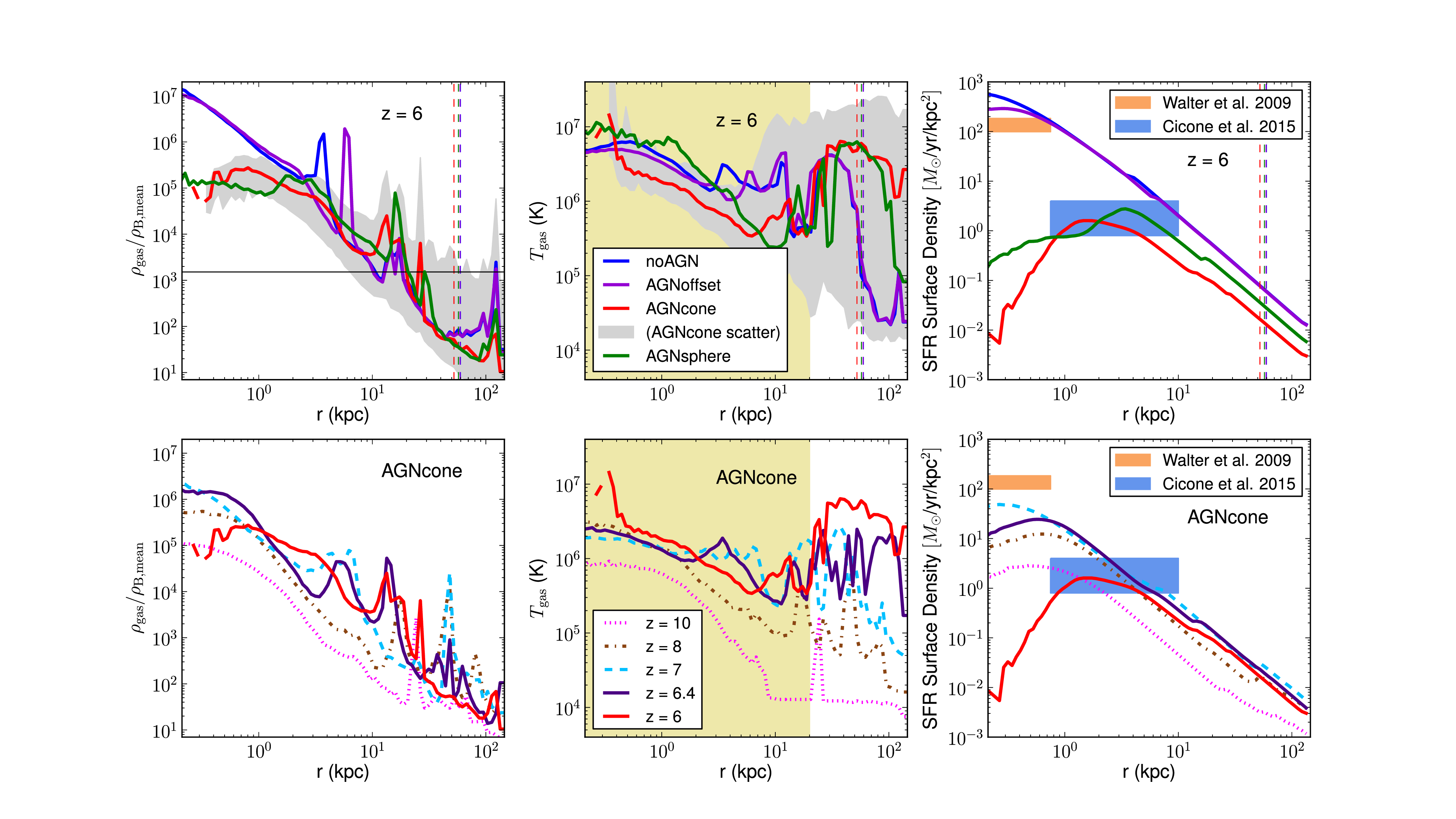} 
\caption{ 
Radial profiles of gas properties around the most-massive galaxy: 
overdensity in the left column, temperature at the middle column, and SFR surface density in the right column. 
Note that the radial coordinate (x-axis) is in units of physical kpc. 
The plotted $\rho$ and $T$ curves denote the median quantity in radial bins of each run. 
The grey shaded area enclose the $90$th percentiles above and below the median of run {\it AGNcone}, 
as a representative of the radial scatter. 
Top row presents the results at $z = 6$, with the colors distinguishing four simulations: 
{\it noAGN} - blue, {\it AGNoffset} - violet, {\it AGNcone} - red, {\it AGNsphere} - green. 
The vertical colored dashed lines in the top panels indicate the halo virial radius ($R_{200}$) in the runs. 
The horizontal black line in the top-left panel denotes the star-formation threshold density 
($n_{\rm SF}$, \S\ref{sec-num-cool-SF-SN}). 
In the $T$ profiles, the gas with a density higher than $n_{\rm SF}$ have been marked with the yellow shaded region; 
since the temperature of such dense gas is not predicted correctly within our SF model. 
Bottom row displays the profiles in a single run {\it AGNcone} at four redshifts: 
$z = 10$ - magenta dotted, $z = 8$ - brown dot-dashed, $z = 7$ - deepskyblue dashed, and $z = 6$ - red solid. 
Observational data is overplotted in the SFR surface density profiles: 
a compact component \citep{Walter09} as the orange shade, 
and an extended component \citep{Cicone15} as the blue shaded area. 
} 
\label{fig-Gas-Rho-Temp-SFR-Profile} 
\end{figure*} 


We plot in Fig.~\ref{fig-Gas-Rho-Temp-SFR-Profile} the 
radial profiles of gas overdensity (left panel), temperature (middle), and SFR surface density (right panel). 
Note that in the $T$ profiles, the gas with $\rho > n_{\rm SF}$ have been marked with the yellow shaded region, 
because the temperature of such dense gas is not predicted correctly, 
but set to the effective equation state describing the SF model \citep[for details see Fig. 1 of ][]{SH03}. 

The top row displays the results of four runs at $z = 6$. 
The density and SFR profiles have similar trends, 
which is expected because star formation in the gas is solely dependent on density 
(our model \S\ref{sec-num-cool-SF-SN}). 
At larger-radii ($r > 10$ kpc), the median density and SFR of all the runs are almost indistinguishable from each other. 
The radial profiles in the run {\it AGNoffset} (violet curve) is similar to that in {\it noAGN} (blue) at all radii, 
because here the BHs do not grow massive enough and there is no efficient feedback. 

In the {\it AGNcone} (red curve) and {\it AGNsphere} (green) runs, where the BHs become supermassive, 
the density and SFR are $10 -100$ times lower at $r \leq 10$ kpc than in the {\it noAGN} run. 
This is due to the central BH, which accretes some gas in and ejects out feedback energy. 
Some central gas is expelled in the form of high-velocity outflows, 
which thermalize their kinetic energy and shock-heat the galaxy outskirt regions. 
This heating is exhibited by the median temperature (top-middle panel) 
remaining high $T \sim 5 \times 10^{6}$ K, in the outer regions at $r \sim 20 - 100$ kpc. 

The bottom row of Fig.~\ref{fig-Gas-Rho-Temp-SFR-Profile} presents the profiles at four redshifts in the run {\it AGNcone}. 
It visualizes the time evolution of the impact of AGN feedback on gas radial properties. 
Initially, the density and consequently SFR profiles increase at all radii 
from $z = 10$ (magenta dotted curve) to $z = 8$ (brown dot-dashed), 
and somewhat further to $z = 7$ (deepskyblue dashed). 
During galaxy formation, gas inflows and collapses to form dense structures where stars form. 
Feedback from the BH acts between $z = 7$ to $z = 6$ (red solid curve), 
when the central (at $r < 5 - 6$ kpc) density drops by a factor $30$ and the central SFR reduces by a factor $10$. 
The temperature at the galaxy outskirts ($r \sim 70 - 700$ kpc) is $T \sim 10^{4}$ K at $z = 10$. 
It increases steadily, by heating from dynamical merger shocks as well as AGN feedback-induced outflows, 
to reach $T \sim 5 \times 10^{6}$ K at $z = 6$. 

Observational data of SFR surface density in the host galaxy of the quasar SDSS J1148 at $z = 6.4$ 
is overplotted in the right panels. 
Our simulations reproduce the SFR surface density of the extended component \citep{Cicone15} 
between $r = 1 - 10$ kpc. 
However the results in the {\it AGNcone} run remains 
below the compact component \citep{Walter09} by a factor $\geq 10$, 
while the runs {\it noAGN} and {\it AGNoffset} are above it by a factor $\sim 2$. 
These imply that the observed quasar SDSS J1148 represents a stage in its evolution 
when star formation is not quenched significantly at the center. 

We summarize that BH feedback decreases gas density at the galaxy center by a factor $100$, 
and increases gas temperature near $R_{200}$ by a factor $50$. 
The former leads to a reduction of SFR surface density by up to a factor $1000$ within $10$ kpc from the galaxy center. 
Powerful BH outflows are launched at epochs between $z = 7$ to $z = 6$, 
which heats up the circumgalactic and intergalactic medium. 


%
%


\subsection{Star-Formation Quenching Mechanism} 
\label{sec-res-Quench} 

\begin{figure*} 
\centering 
\includegraphics[width = 1.13 \linewidth]{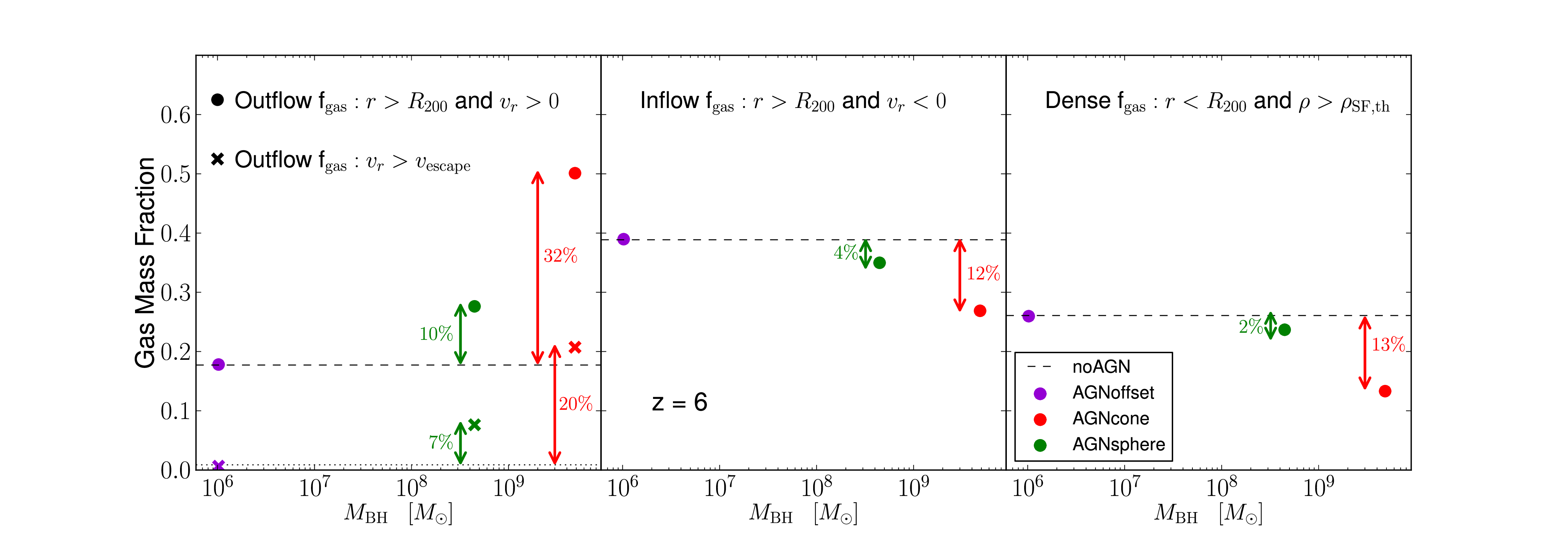} 
\caption{ 
Mass fraction of outflowing, inflowing and dense gas as a function of BH mass, at $z = 6$. 
Left panel shows outflowing gas fraction computed in two ways: 
gas with radial velocity larger than the halo escape velocity $(v_r  > v_{\rm escape})$ as the cross symbols, 
and outgoing gas outside the halo virial radius $(r > R_{200}$ and $v_r  > 0)$ as the filled circles. 
Middle panel shows inflowing gas fraction outside the virial radius $(r > R_{200}$ and $v_r < 0)$. 
Right panel shows gas fraction denser than the star-formation threshold density 
inside the virial radius $(r < R_{200}$ and $\rho > \rho_{\rm SF, th})$. 
Each of the three points is for one AGN simulation, as distinguished by the color. 
The horizontal black-dashed line (and the black-dotted line in the left panel) 
indicates the corresponding fraction in the {\it noAGN} run. 
The fraction in each simulation is computed with respect to all the gas within the zoomed-in volume. 
} 
\label{fig-Frac-OutFlow-vs-BHmass} 
\end{figure*} 

The fraction of outflowing, inflowing and dense gas 
as a function of BH mass at $z = 6$ is plotted in Fig.~\ref{fig-Frac-OutFlow-vs-BHmass}. 
The mass fraction in each simulation is computed with respect to all the gas within the zoomed-in region. 
The left panel shows outflowing gas fraction computed using two conditions: 
gas with radial velocity larger than the halo escape velocity $(v_r  > v_{\rm escape})$ as the cross symbols, 
and outgoing gas outside the halo virial radius $(r > R_{200}$ and $v_r  > 0)$ as the filled circles. 
The middle panel shows inflowing gas fraction outside the virial radius $(r > R_{200}$ and $v_r < 0)$. 
The right panel shows gas fraction denser than the star-formation threshold density 
inside the virial radius $(r < R_{200}$ and $\rho > \rho_{\rm SF, th})$. 

Similar to the other results in run {\it AGNoffset} (violet point), where the BHs do not grow, 
the gas fractions are the same as that in the {\it noAGN} case 
(horizontal black-dashed line in all panels, and the black-dotted line in the left panel). 
Effects of BH feedback and the subsequent outflows created can be seen 
in runs {\it AGNcone} (red) and {\it AGNsphere} (green) as the higher outflowing fractions, 
lower inflowing fraction, and lower dense fraction. 
We find a positive correlation of the outflowing gas mass fraction with the central BH mass in galaxies. 

The differences caused by the most-massive BH, compared to the SF-only case, 
is written as the percentage value in each panel. 
A central BH of $M_{\rm BH} = 4 \times 10^9 M_{\odot}$ unbinds $20 \%$ of the gas 
in and around its host galaxy, which outflows with $v_r  > v_{\rm escape}$ and escapes the halo. 
It makes $32 \%$ more gas outside $R_{200}$ to outflow. 
It reduces the inflow fraction outside $R_{200}$ by $12 \%$. 
It lowers the high-density (denser than $\rho_{\rm SF, th}$) gas fraction inside $R_{200}$ by $13 \%$. 
All these processes limit the gas available for star-formation. 

Our results are consistent with recent hydrodynamical simulation analyses by \citet{Beckmann17}, 
who concluded that SMBHs affect their host galaxies through a combination of outflows 
and disruption of central gas inflows, which acting together drops net inflows by up to $70\%$. 

We deduce that AGN feedback effects quench star-formation in galaxies by a combination of mechanisms. 
The cosmic inflow fraction lowered is comparable to the dense gas fraction reduced. 
Fast outflowing gas ejected away affects twice the fraction of gas as compared to the two above, 
which is hence the dominant process. 


\section{Summary and Conclusions} 
\label{sec-conclusion} 

Quasars are observed with supermassive black holes at their centers, 
some hosting powerful outflows, even at high-redshifts. 
We investigate the impact of AGN feedback in massive galaxies in the early universe, 
and probe $z \geq 6$ quasar outflows, using numerical simulations. 
We employ a modified version of the SPH code GADGET-3. 
It includes the following sub-resolution physics: radiative cooling and heating from photoionizing background, 
star-formation, stellar evolution, chemical enrichment for $11$ elements, 
supernova feedback, AGN accretion and feedback. 
We use novel methods to distribute the feedback energy from a BH in the kinetic form, 
where the velocity of the neighbouring gas is incremented. 

We perform zoomed-in cosmological hydrodynamical simulations of quasar-host galaxies. 
A $(500 ~ {\rm Mpc})^3$ comoving volume is first evolved from $z = 100$ using $256^3$ dark-matter particles. 
We select the most-massive halo at $z = 6$ of total mass $4.4 \times 10^{12} M_{\odot}$, 
and zoom-in a cubic region of side $2 R_{200} \simeq 1022$ kpc comoving. 
The Lagrangian volume of $(5.21 ~ {\rm Mpc})^3$, tracked back to our initial condition at $z = 100$, 
is populated with high-resolution DM and baryon particles. 
The gas particle mass is $1.41 \times 10^{6} M_{\odot}$, and the gravitational softening length is $1 /h$ kpc comoving. 
The zoom-in simulation is finally evolved from $z=100$ up to $z = 6$, 
containing high-resolution particles in the $(5.21 ~ {\rm Mpc})^3$ volume, 
embedded inside the $(500 ~ {\rm Mpc})^3$ low-resolution periodic box. 


We execute four runs: one of them is a control simulation with SF-only and no BH; 
the other three runs are with AGN feedback exploring different sub-resolution models for the BHs. 
In our simulation input model, 
we seed BHs of mass $10^5 M_{\odot}$ at the centers of massive halos with $M_{\rm halo} > 10^{9} M_{\odot}$. 
The earliest BHs appear at $z \sim 15$. 
The BHs are allowed to grow by accreting surrounding gas and by merger with other BHs. 
As they accrete and grow, the BHs eject out feedback energy. 
We analyze the simulations by exploring the growth of the first SMBHs, 
their coevolution with galaxies, and the generated outflows. 
We find the following results from our simulations: 

\begin{itemize} 

\item The BHs grow to supermassive ($M_{\rm BH} \sim 10^9 M_{\odot}$) 
only when we implement the repositioning or advection model, where a BH is repositioned to the center of its host galaxy. 
The BHs then accrete gas at the Eddington accretion rate over $z = 9 - 6$, 
by encountering galaxy-central high-density gas. 
The most-massive BH at $z = 6$ has: 
$M_{\rm BH} = 4 \times 10^9 M_{\odot}$ and $\dot{M}_{\rm BH} = 100 M_{\odot}$/yr. 

\item Massive BHs at galaxy centers mostly occur as merging systems at $z = 6 - 7$, 
and leads to the formation of a SMBH binary or triplet. 
This prediction makes a strong case to observe them at high-$z$ in the future. 

\item The BHs are too small ($M_{\rm BH} \sim 10^6 M_{\odot}$) in the case without BH repositioning. 
The growth is dominated by mergers, since gas density and consequently accretion rate is always low. 
The results here are similar to that in the noAGN case, because there is no effective AGN feedback. 

\end{itemize} 

We see the following AGN feedback impacts when the BHs grow supermassive $M_{\rm BH} \geq 10^8 - 10^9 M_{\odot}$. 

\begin{itemize} 

\item Star-formation is quenched between $z = 8 - 6$, when the BHs have reached $M_{\rm BH} > 10^7 M_{\odot}$. 
The total SFR is reduced by a factor $5$ (to $200 M_{\odot}$/yr), 
and the SFR surface density lowered by $1000$ near the galaxy center. 

\item The SMBHs in massive galaxies $(M_{\star} > 10^{9} M_{\odot})$ at $z = 8 - 6$ are 
more-massive than expected from the local $[M_{\rm BH} - M_{\star}]$ correlation; 
and the same scenario is indicated by observations at $z \geq 6$. 


\item Bipolar bubble-like high-velocity ($v_{r} \sim 1400 - 2000$ km/s) outflows form 
at epochs $z = 7$ to $z = 6$, originating from the BH. 
The outflows are hot, consist of shock-heated low-density gas, 
and propagate beyond the galaxy virial radius, $R_{\rm 200}$ (i.e. up to a few hundreds kpc). 

\item Cold dense filamentary cosmic gas inflows are disrupted by the AGN outflow-driven forward shocks, 
and reduced by a factor $30 \%$ near $r=R_{200}$ (and by $12 \%$ considering the gas outside $R_{200}$) at $z = 6$. 
This halts the formation of satellites, thus altering the substructure distribution in the inner parts of the galaxy. 
While at the same time, significant amounts of gas ($\sim 30 - 40 \%$) continue to inflow towards the central region, 
perpendicular to the outflow direction. 

\item The density of gas is decreased at the galaxy center ($r < 10$ kpc) by a factor $100$. 
The temperature is increased at the outskirts ($r \sim 100 - 800$ kpc) by a factor $50$. 

\item A fraction $\sim 20 \%$ of the total gas outflows with a velocity larger than the escape speed, 
thus being able to escape the galaxy. 
While in the noAGN case, no gas can escape from the galaxy. 
There is a positive correlation between the outflowing gas fraction and the central BH mass. 

\end{itemize} 

{\it We deduce that AGN feedback quenches star-formation in galaxies by a combination of mechanisms: 
ejecting gas out of the galaxy halo at high-velocity, reducing the amount of dense gas, as well as halting gas inflows.} 
A BH of $M_{\rm BH} = 4 \times 10^9 M_{\odot}$ drives 
$32 \%$ more gas outside $R_{200}$ as outflows (with $20 \%$ larger fraction above the $v_{\rm escape}$), 
reduces the inflow fraction outside $R_{200}$ by $12 \%$, 
and lowers the high-density (denser than $\rho_{\rm SF, th}$) gas fraction inside $R_{200}$ by $13 \%$. 
All these processes limit the gas available for star-formation. 
{\it The dominant contribution to this effect comes from fast, powerful outflows ejecting gas out of the host galaxy halo.} 


\section*{Acknowledgments} 

We thank the referee for useful comments which helped to clarify the paper. 
We are most grateful to Volker Springel for allowing us to use the GADGET-3 code. 
We thank Giuseppe Murante, Klaus Dolag, and Alexander Beck for technical help with the code. 
This work is supported by the PRIN-INAF 2014 grant ``Windy black holes combing galaxy evolution". 
CC acknowledges funding from the European Union's Horizon 2020 research and innovation programme 
under the Marie Sklodowska-Curie grant agreement No 664931. 
RM acknowledge support from the Science and Technology Facilities Council (STFC) 
and from the ERC Advanced Grant 695671 ``QUENCH". 
SC acknowledges financial support from the STFC.

%



\appendix 

\section{} 
\label{sec-app} 

\begin{figure*} 
\centering 
\includegraphics[width = 1.1 \linewidth]{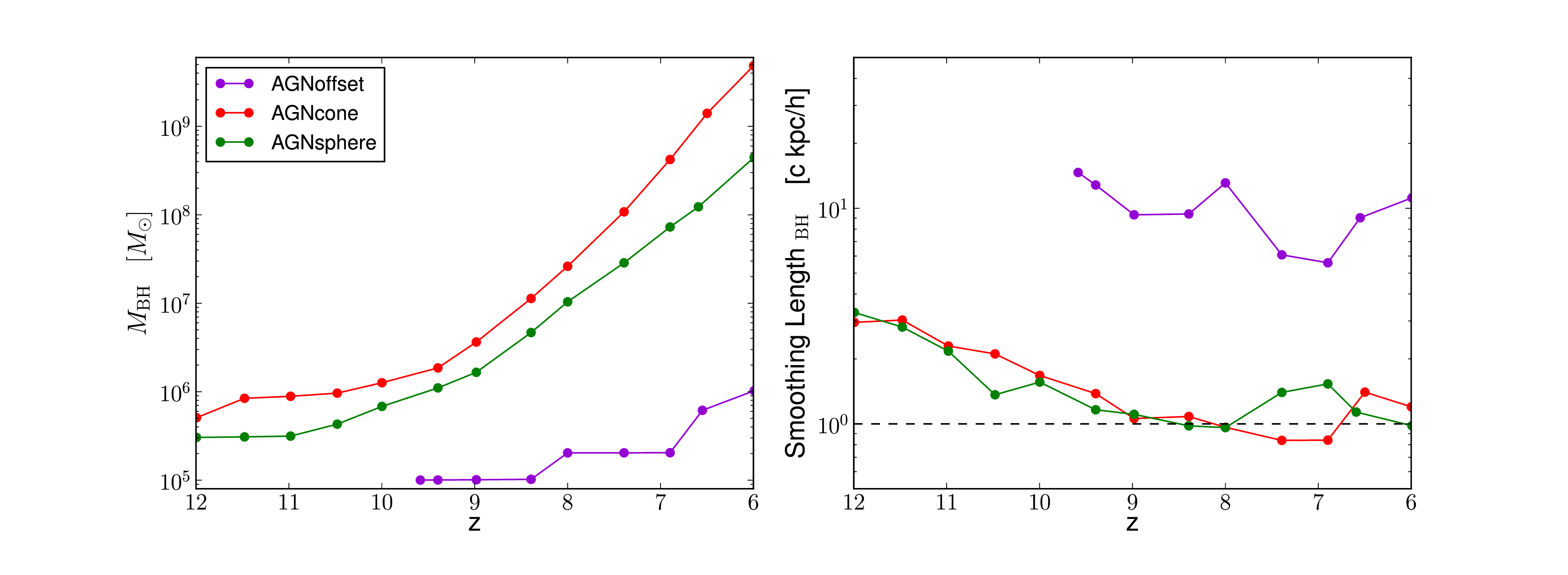} 
\vspace{-1.7cm} \\ 
\caption{ 
Evolution with redshift of BH mass (left panel) and BH smoothing length (right) of the most-massive BH in each run. 
These are the same BHs which were plotted in Fig.~\ref{fig-BH-Mass-AccrRate-SFRD}. 
The different colors discriminate the runs as labelled in the left panel. 
The horizontal dashed line in the right panel indicate the gravitational softening length 
for the high-resolution DM and gas particles. 
} 
\label{fig-BH-Mass-hSml} 
\end{figure*} 

In our implementation of BH kinetic feedback (\S\ref{sec-num-BH-Accr-Feed}), 
all the $\sim 200$ neighboring gas particles lying within the bi-cone or sphere 
have the same probability to be kicked (Eq.~\ref{eq-probKick}). 
This could potentially cause a problem as gas particles can be kicked and AGN wind can be launched far from the BH itself. 
The problem will cause the BH to clear out gas around it, 
and the radius of influence of the BH (its smoothing length) would grow artificially large. 

We checked that there is no unphysical behavior of the BH smoothing length ($h_{\rm BH}$) 
because of our adopted numerical prescription. 
Therefore the potential problem does not occur. 
The number of gas particles actually kicked by a BH at a timestep is just a few (maximum $5$ to $6$) in our simulations. 

The redshift evolution of the BH smoothing length (Eq.~\ref{eq-BH-Smooth}) 
of the most-massive BH in the three AGN runs is plotted in Fig.~\ref{fig-BH-Mass-hSml}, right panel. 
These are the same BHs whose mass vs. redshift are plotted in Fig.~\ref{fig-BH-Mass-AccrRate-SFRD}. 
For comparison, the horizontal dashed line indicate the gravitational softening length 
($L_{\rm soft} = 1 /h$ kpc comoving) for the high-resolution DM and gas particles. 

Fig.~\ref{fig-BH-Mass-hSml} demonstrates that $h_{\rm BH}$ depends dominantly (and inversely) 
on the density of gas where the BH lies. 
Gas density at galaxy centers becomes higher with time. 
Therefore, as galaxy-central BHs grow more massive with time, they have a smaller $h_{\rm BH}$. 
There is no abnormal increase of $h_{\rm BH}$, which looks to have an expected evolution. 
We see that $h_{\rm BH} \sim L_{\rm soft}$ for most of the time, 
over redshifts $z = 9 - 6$, when the BHs have the highest growth. 

At $z > 9$, $h_{\rm BH}$ of the galaxy-central BHs goes up to a maximum $\sim 3 L_{\rm soft}$. 
At such early epochs, there is the possibility to apply a kind of distance-weighted, or density-dependent 
expression for the probability that a gas particle is kicked 
({\it such that the closest gas particles having the highest probability to be ejected}). 
Such alternative numerical implementation are however beyond the goals of the current work, 
and deferred to future studies. 
At the present state, it is unclear if and how the choice of distance and density dependence 
of the kicking probability can affect the resulting feedback. 
Further (and preferably higher-resolution) simulations are required to perform 
a dedicated study on the numerical aspects. 





\end{document}